\documentstyle[eqsecnum,aps,prb,epsfig]{revtex}

\begin{document}
\draft \preprint{
}
\title{ Binary self-similar one-dimensional quasilattices:\\ Mutual
local-derivability classification and substitution rules
}
\author{Masashi Torikai } \address{Department of Physics, Kyushu
University, Fukuoka 812-8581 }
\author{Komajiro Niizeki }
\address{Department of Physics, Graduate School of Science, Tohoku
University, Sendai 980-8578 }
\author{Takashi Odagaki }
\address{Department of Physics, Kyushu University, Fukuoka 812-8581 }
\maketitle

\begin{abstract}
Self-similar binary one-dimensional (1D) quasilattices (QLs) are
classified into mutual local-derivability (MLD) classes. It is shown
that the MLD classification is closely related to the
number-theoretical classification of parameters which specify the
self-similar binary 1D QLs. An algorithm to derive an explicit
substitution rule, which prescribes the transformation of a QL into
another QL in the same MLD class, is presented. An explicit inflation
rule, which prescribes the transformation of the self-similar 1D QL
into itself, is obtained as a composition of the explicit substitution
rules. Symmetric substitution rules and symmetric inflation rules are
extensively discussed.
\end{abstract}

\section{Introduction}\label{sec_introduction}

 A quasiperiodic structure lacks the periodicity but has a long-range
 positional order. Quasiperiodic structures have been intensively
 investigated since the experimental discovery of quasicrystals, which
 have quasiperiodic structures with non-crystallographic rotational
 symmetries.~\cite{Ya96} In order to discuss the peculiar properties of
 quasicrystals, many models for the structures of quasicrystals have
 been proposed. The most successful model is based on a quasilattice
 (QL), which is a quasiperiodic and discrete set of points.
 
 The mutual local-derivability (MLD) is one of the most important
classification schemes for QLs.~\cite{cite1,cite2} According to a recent
study of a tight-binding model on a one-dimensional (1D) QL, two 1D QLs
belonging to different MLD classes belong to different universality
classes with respect to their one-electron properties.~\cite{cite3}

 Some of the quasiperiodic structures have self-similarities. In
 particular, all the quasicrystals found experimentally are closely
 related to self-similar quasiperiodic structures.~\cite{Ya96,cite4} On
 the other hand, the self-similarity has been used successfully for a
 rigorous study of the one-electron properties of 1D QLs because it
 matches well renormalization methods such as the trace
 map.~\cite{cite5,cite6,cite7,cite8}

 Many studies have focused on self-similar 1D quasiperiodic structures,
 but some important points have been left.  Kaneko and Odagaki have
 discussed the quasiperiodic sequence of letters (i.e., objects without
 geometrical length), which are produced by associating different
 letters to different types of intervals (atomic distance) in
 QLs.~\cite{s2,s3} They have shown a procedure to obtain an explicit
 inflation rule (IR), which prescribes the transformation of a
 self-similar quasiperiodic structure into itself, for a binary
 quasiperiodic sequence of letters characterized by a quadratic
 irrational. However, since the quasiperiodic sequence of letters is not
 a geometrical object, the geometrical meaning of the explicit IR is
 obscure.  Binary sequences derived from substitution rules (IRs in our
 terminology) are extensively investigated in Ref.\ \onlinecite{cite7}.
 They found that there exist hierarchical series of substitution rules
 which have good properties with respect to the trace map.  They
 discussed interrelations among different substitution rules.  This work
 also treats the binary sequences as sequences of letters. Yamashita has
 investigated the relation between the projection method and two series
 of binary quasiperiodic sequences of letters with inflation
 symmetries.~\cite{Ya93} However, he has not treated the sequences as
 geometrical objects.  As a consequence of considering the quasiperiodic
 sequence of letters instead of the QL itself, these works have not
 revealed the geometrical meaning of the self-similarity and the IR .

 In order to consider the geometrical meaning of the self-similarity and
 the IR, it is necessary to investigate the QLs instead of the sequences
 of letters. In Ref.\ \onlinecite{s1}, the self-similarity of 1D QLs has
 been investigated, and the geometrical meaning of the self-similarity
 has been revealed.  However, a general algorithm to obtain an explicit
 IR has not been shown. In Ref.\ \onlinecite{NF}, the inflation symmetry
 of binary and ternary 1D QLs has been extensively investigated but this
 study has been restricted to the cases where the quadratic irrationals
 characterizing the QLs belong to two infinite series, one of which is
 the series composed of all the quadratic irrationals of rank one in our
 terminology to be presented in a later section, while the other is a
 subseries of quadratic irrationals of rank two.

 In the present paper, we restrict our considerations to self-similar
 binary 1D QLs but we do not impose any condition on the quadratic
 irrationals characterizing the QLs. The purpose of the present paper is
 to classify all the self-similar binary 1D QLs into MLD classes and to
 derive explicit {\it substitution rules} (SR), which relate different
 members of a single MLD class.

 In the next section, we make a review of the self-similarity of binary
 QLs. Here we identify parameters which characterize these QLs. In Sec.\
 \ref{sec_mld}, we classify the QLs into MLD classes on the basis of the
 number-theoretical properties of the parameters. Also, we give an
 explicit SR which combines two QLs in a single MLD class. In Sec.\
 \ref{sec:composite}, we discuss composite SRs and IRs. We show several
 examples of self-similar binary QLs in Sec.\ \ref{sec_examples}. In
 Sec.\ \ref{sec:symmetric}, we discuss symmetric SRs and symmetric
 IRs. Miscellaneous subjects are discussed in Sec.\
 \ref{sec:miscellanious}. In the last section, we summarize the results,
 and discuss a few remained subjects.
 
 In order to develop our theory, several results and technical terms in
the number theory of quadratic irrationals are necessary, and they are
explained in Appendix \ref{glossary}. We will sometimes use them in
later sections without explicit references to the appendix.  We suggest
the readers jumping to Appendix at this point although it is
considerably long.

 Finally, a list of some of abbreviated words appearing in the present
paper is presented: {\it inflation rule} (IR), {\it substitution rule}
(SR), {\it mutual local-derivability} (MLD), {\it unimodular Frobenius
matrix} (UFM), {\it modular equivalent class} (MEC), {\it
quasi-permutation-matrix} (QPM), {\it reduced quadratic irrational}
(RQI), and {\it quasi-reduced quadratic irrational} (QRQI). The last
five words are defined and discussed in Appendix.

\section{Self-Similarity of Binary 1D QLs} \label{sec_ss}

\subsection{Binary QLs in the projection method} \label{Binary QLs}

\begin{figure}
  \begin{center}
   \centerline{\epsfig{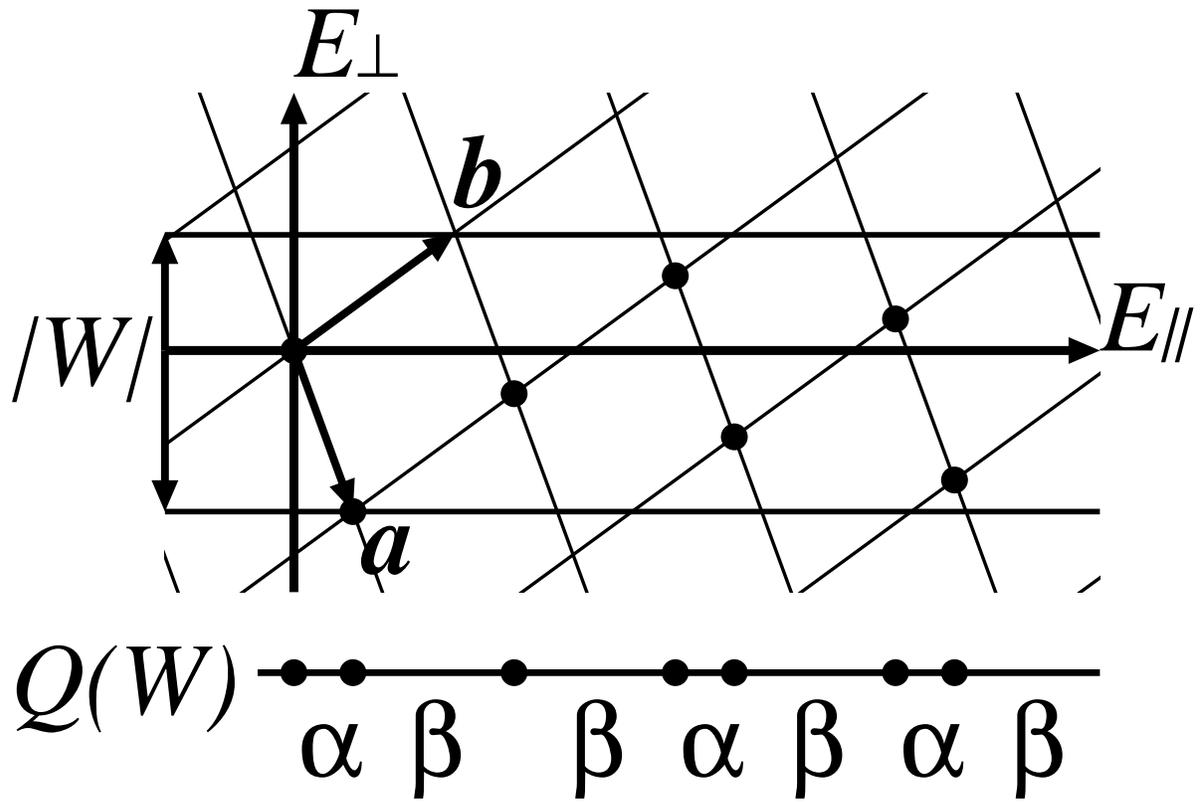}}
   \caption{A derivation of a binary 1D QL by the projection method. The
   mother lattice $\Lambda$ is cut with the strip, and then projected
   onto $E_{\parallel}$.} \label{fig1}
   \end{center}
 \end{figure}

 A QL is derived, as shown in Fig.~\ref{fig1}, with the projection
 method from a 2D periodic lattice $\Lambda$, which is called a mother
 lattice.
 The 2D space embedding $\Lambda$ is spanned by two orthogonal
 subspaces $E_{\parallel}$ and $E_{\perp}$, which are called the
 physical space and the internal space, respectively. We shall use
 throughout this paper the coordinate system in which the $x$ axis and
 the $y$ axis are chosen to be $E_{\parallel}$ and $E_{\perp}$,
 respectively. The physical space is assumed to be in an incommensurate
 configuration with $\Lambda$. One of the two basis vectors, ${\bf a}$
 and ${\bf b}$, of $\Lambda$ is assumed to be in the first quadrant and
 the other in the fourth quadrant, so that the fundamental parallelogram
 which is spanned by ${\bf a}$ and ${\bf b}$ is cut by
 $E_{\parallel}$. If all the lattice points within a strip which is
 parallel to $E_{\parallel}$ are projected onto $E_{\parallel}$, we
 obtain a QL as shown in Fig.~\ref{fig1}. The QL divides naturally
 $E_{\parallel}$ into intervals, so that we shall identify the QL with
 the relevant 1D tiling of $E_{\parallel}$. In the present paper, we
 restrict our considerations to binary QLs.

 We call the section of the strip cut by $E_{\perp}$ a window.  A binary
 QL is obtained as shown in Fig.~\ref{fig1} if the size of the window is
 equal to the vertical width of the fundamental parallelogram.  The two
 types of intervals are denoted as $\alpha$ and $\beta$, whose lengths,
 $a$ and $b$, are equal to the $x$-components of the two basis vectors.
 It is essential in obtaining a binary QL that one of the two basis
 vectors of $\Lambda$ is in the first quadrant and the other in the
 fourth quadrant and that the size of the window is equal to the
 vertical width of the fundamental parallelogram.  Therefore, we shall
 call such a set of basis vectors a canonical set and the relevant
 window a canonical window; a canonical window is associated with a
 canonical set.  Since there exist an infinite variety of canonical sets
 of basis vectors, there exist an infinite variety of canonical windows.
 This does not mean, however, that there exist an infinite variety of
 binary QLs with a common mother lattice because different QLs can be
 (geometrically) similar to one another.  It will be shown in a later
 section that, under a certain condition, there exist only a finite
 number of binary QLs which are derived from a common mother lattice but
 not mutually similar; the condition is that the mother lattice has the
 {\it hyperscaling symmetry}.~\cite{NF}

 Two QLs are called to be mutually {\it locally isomorphic} (LI) if
 every finite part of one of them is always found in the other and vice
 versa; two QLs being mutually LI cannot be distinguished
 macroscopically from one another. A necessary and sufficient condition
 for two QLs to be mutually LI is that the relevant two windows have a
 common size. In the present paper, we will not distinguish two QLs if
 they are mutually LI. We denote $Q \simeq Q'$ when two QLs, $Q$ and
 $Q'$, are mutually LI. A QL derived from a mother lattice is uniquely
 determined by the window, $W$, and the QL specified by $W$ is denoted
 as $Q(W)$. We will not hereafter distinguish the window, $W$, and its
 size, $|W|$.

 It is known that a Fourier module, i.e., an additive group formed of
reciprocal lattice vectors of a QL, is given by projecting the Fourier
module of the mother lattice onto the reciprocal physical
space. Therefore different QLs derived from a single mother lattice but
with windows of different sizes have a common Fourier module.

The scale of the internal space, $E_\perp$, is irrelevant on the
construction of a QL because the projection is made along $E_\perp$ in
the projection method. Hence, two mother lattices which only differ in
the scale of the internal space are considered to be isomorphic, and are
not mutually distinguished. On the other hand, a scale change of the
physical space, $E_\parallel$, results in a scaling of a QL, so that to
fix the scale is equivalent to fix the lattice constant of the QL. In a
usual classification scheme of QLs, two QLs being geometrically similar
to each other are assumed to be identical. Then, two mother lattices
which differ in the scales of the physical space and/or the internal
space are sometimes assumed to be identical.

\subsection{Self-similar binary QLs} \label{Self-Similar QLs}

 The Fibonacci lattice is a representative self-similar binary QL. The
sizes of two types of intervals, $\alpha$ and $\beta$, of the Fibonacci
lattice are equal to $1$ and $\tau_{\rm G}$, respectively, where
$\tau_{\rm G}$ is the golden ratio: $\tau_{\rm G} := (1+\sqrt{5})/2$. If
we substitute the intervals $\beta$ and $\alpha\beta$ for $\alpha$ and
$\beta$ in the Fibonacci lattice, respectively, we obtain again the
Fibonacci lattice. This transformation of the Fibonacci lattice into
itself is known as {\it the inflation rule} (IR), which is represented
as
 \begin{equation}
 \alpha \rightarrow \beta, \quad \beta \rightarrow \alpha\beta.
\label{inflation_fibonacci}
 \end{equation}
 Each interval is scaled up by $\tau_{\rm G}$ on the inflation.
 
 The self-similarity of a QL is represented generally by an IR which
prescribes how the inflated versions of the two intervals are divided
into the original intervals.~\cite{s1} The IR accompanies naturally the
inflation matrix $M$, which is a unimodular Frobenius matrix (UFM) as
defined in Appendix \ref{Frobenius}. For example, the inflation matrix
of the Fibonacci lattice is given by
 \begin{equation}
 M = \left( \begin{array}{cc} 0 & 1 \\ 1 & 1
 \end{array} \right). \label{M}
 \end{equation}
 The golden ratio, $\tau_{\rm G}$, is not only the Frobenius eigenvalue
of $M$ but also the companion quadratic irrational of $M$. Since the
explicit form of the inflation matrix $M$ of a self-similar binary QL
depends on the order of the two intervals in the defining equation for
$M$, we shall fix the order so that the first interval is the shorter of
the two. Then, $M$ is a quasi-normal UFM as defined in Appendix
\ref{Frobenius}. A self-similarity of a QL is derived in the projection
method only when the mother lattice has the hyperscaling symmetry: this
symmetry is an automorphism of $\Lambda$ and leaves the physical space
and the internal space invariant.~\cite{s1,NF,cite9}
 
 Since the inflation matrix $M$ is unimodular, it determines a 2D
lattice $\Lambda$ with the hyperscaling symmetry as shown in Appendix
\ref{unimodular matrix}. The physical space (or the internal space) is
chosen to coincide with the expansive (or contractive) principal axis of
the linear transformation $T$ associated with $M$ (see Appendix
\ref{unimodular matrix}). More precisely, $T$ enlarges along
$E_{\parallel}$ by $\tau$, i.e., the Frobenius eigenvalue of $M$, but
shrinks along $E_{\perp}$ by ${\tau}^{-1}=|\bar{\tau}|$. The left
Frobenius eigenvector $(a \;\; b)$ with $a, ~ b > 0$ satisfies the
equation
 \begin{equation}
 \tau(a \;\; b) = (a \;\; b)M, \label{self-similar3}
 \end{equation}
 while the second left eigenvector, $(a_{\perp} \;\; b_{\perp})$,
satisfies the equation
 \begin{equation}
 \bar{\tau}(a_{\perp} \;\; b_{\perp}) = (a_{\perp} \;\; b_{\perp})M.
\label{self-similar4}
 \end{equation}
 The two basis vectors of $\Lambda$ are given by
 \begin{equation}
 {\bf a} = \left(
    \begin{array}{l}
     a \\ a_{\perp}
    \end{array}
  \right), \quad {\bf b} = \left(
    \begin{array}{l}
     b \\ b_{\perp}
    \end{array}
  \right). \label{bases}
 \end{equation}
 It follows from Eqs.\ (\ref{self-similar3}) and (\ref{self-similar4})
that $T$ is an automorphism of $\Lambda$; $T$ yields the hyperscaling
symmetry of $\Lambda$. Since the row vector $(a_{\perp} \;\; b_{\perp})$
is not a Frobenius eigenvector, $a_{\perp}$ and $b_{\perp}$ must have
different signs. We can assume that $a_{\perp} < 0$ and $b_{\perp} > 0$. 
Then, ${\bf a}$ and ${\bf b}$ are in the fourth quadrant and the first
quadrant, respectively, so that $\{{\bf a}, ~{\bf b}\}$ is a canonical
set. The QL $Q(W)$ derived from $\Lambda$ is binary if
  \begin{equation}
    W = - a_{\perp} + b_{\perp}\; (= |a_{\perp}| + b_{\perp}),
   \label{window}
 \end{equation}
 and the two types of intervals of the binary QL have the lengths $a$
and $b$. Eq.\ (\ref{self-similar3}) prescribes how the inflated versions
of the two intervals are divided into the original intervals.
 
 Let us introduce two positive parameters by
 \begin{equation}
    \xi := b/a, \quad \zeta := -a_{\perp }/b_{\perp}. \label{ratio}
 \end{equation}
 Then, Eqs.\ (\ref{self-similar3}) and (\ref{self-similar4}) are
rewritten as
 \begin{equation}
  \tau (1 \;\; \xi)\;\;\;=(1 \;\; \xi)M, \label{left eigenvector1}
 \end{equation}
 \begin{equation}
   \bar{\tau} (-\zeta \;\; 1)=(-\zeta \;\; 1)M. \label{left
   eigenvector2}
 \end{equation}
 Note that $\xi > 1$ because $a < b$. Since $M$ is an integer matrix,
the row vector $(1 \;\; {\bar \xi})$ must be also a left eigenvector
corresponding to the eigenvalue $\bar{\tau}$ of $M$. Hence we have
$\zeta = {\tilde \xi}$ with ${\tilde \xi} := -1/{\bar \xi}$ being the
dual to $\xi$ as defined in Appendix \ref{unimodular matrix}. It can be
shown that the ratio of the frequency of occurrence of the interval
$\alpha$ to that of $\beta$ is equal to $\zeta$. Although $\xi$ and
$\zeta$ belong both to the quadratic field ${\bf Q}[\tau]$, their
dependences on $\tau$ are different in general from the case of the
Fibonacci lattice for which $\xi = \zeta = \tau \;(=\tau_{\rm G})$.
 
 The projections of $\Lambda$ onto $E_{\parallel}$ and $E_{\perp}$ yield
modules, $\Lambda_{\parallel}$ and $\Lambda_{\perp}$, which are dense 1D
sets. They are isomorphic to $\Lambda$, and there are a natural
bijection between any pair of the trinity, $\{\Lambda, ~
\Lambda_{\parallel}, ~\Lambda_{\perp}\}$.~\cite{cite9} A QL is nothing
but a discrete subset of $\Lambda_{\parallel}$. The projection modules
are equal to the ${\bf Z}$-modules ${\bf Z}[\xi]$ and ${\bf Z}[\zeta]
\;(= {\bf Z}[{\tilde \xi}])$, respectively, where the physical space and
the internal space are scaled so that $a = b_{\perp} = 1$, which we will
sometimes assume. Remember that $p + q\xi \in {\bf Z}[\xi]$ and $-p\zeta
+ q \in {\bf Z}[\zeta]$ are combined by the bijection mentioned above.

 A well-known formula for a binary QL yields the position of the $n$-th
site (lattice point) of $Q(W)$ with $W = 1 + \zeta$:
 \begin{equation}
    x_n = n + (\xi - 1) \left\lceil\frac{n + \phi}{1 +
\zeta}\right\rceil, \label{site}
 \end{equation}
where $\phi$ is the phase parameter depending on the position of the
window in the internal space.
 
\begin{figure}
  \begin{center}
  \centerline{\epsfig{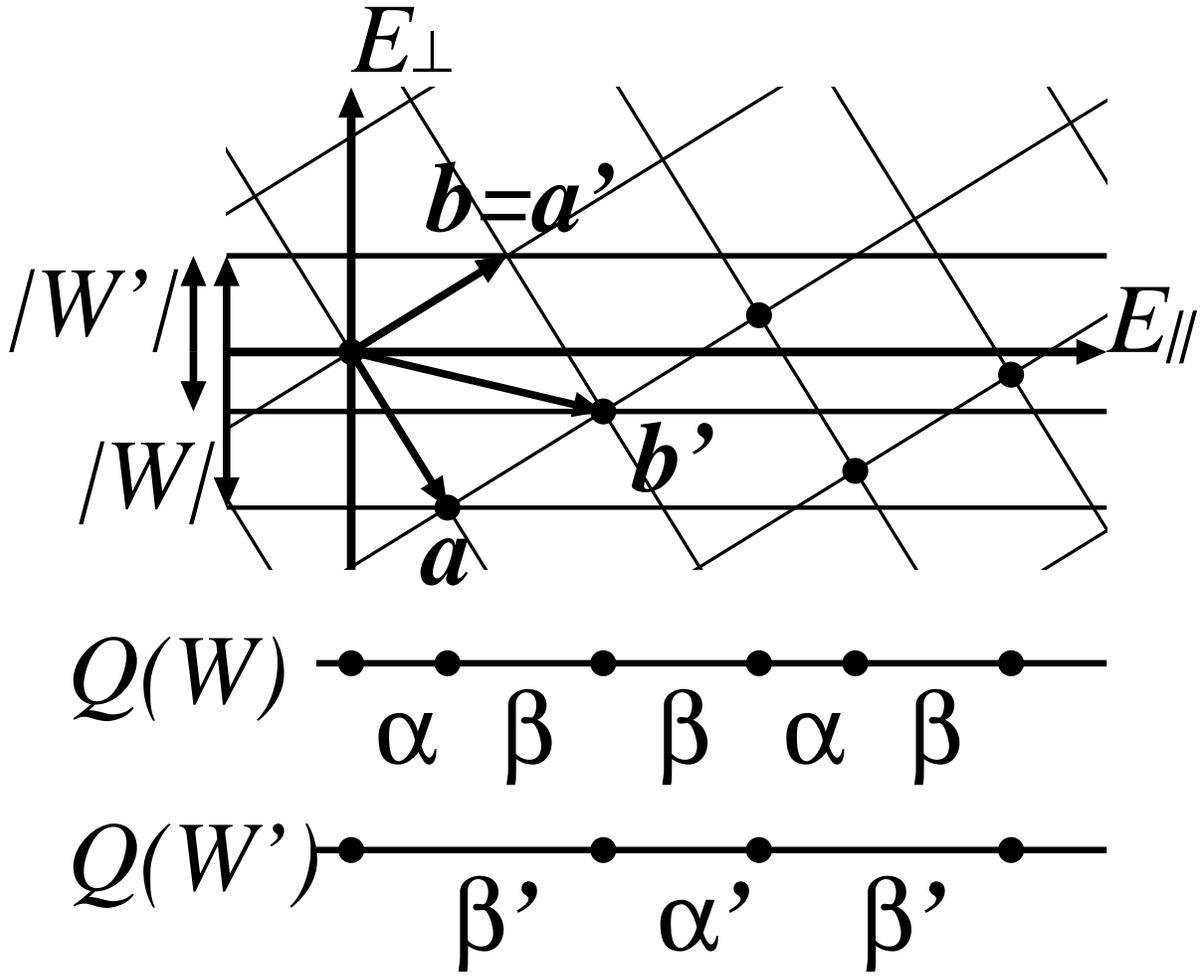}}
  \caption{A geometrical derivation of the IR of the Fibonacci lattice. 
  The canonical set of basis vectors, $\{{\bf a}, ~{\bf b}\}$, is
  transformed by the hyperscaling operation $T$ into another set,
  $\{{\bf a}', ~{\bf b}'\}$. The two types of new intervals, $\alpha '$
  and $\beta '$, are related to the original intervals, $\alpha $ and
  $\beta $, by the IR. The window is shrunken by $T$ from $W$ to
  $W'={\tau}^{-1}W$, while the sizes of the two types of intervals are
  expanded by the factor $\tau$.} \label{fig2}
   \end{center}
 \end{figure}

 It is important that the hyperscaling operation $T$ transforms a
canonical set, $\{{\bf a}, ~{\bf b}\}$, into another canonical set,
$\{{\bf a}', ~{\bf b}'\} := \{T{\bf a}, ~T{\bf b}\}$. The new canonical
set yields another binary QL, $Q(W')$, where the new window $W'$ is
related to the old one by $W'={\tau}^{-1}W$. The sizes of the two types
of intervals $\alpha '$ or $\beta '$ of $Q(W')$ are given by $a' := \tau
a$ and $b' := \tau b$. The hyperscaling symmetry of $\Lambda$ results in
that $\tau Q(W) \simeq Q(W')$, which means that $Q(W)$ is a self-similar
QL, whose ratio of the self-similarity is equal to $\tau$.~\cite{NF}
However, it is not obvious whether the two types of intervals $\alpha '$
and $\beta '$ are uniformly decorated by the two types of intervals
$\alpha$ and $\beta$ because the inflation matrix does not directly
prescribe the order of the latter intervals in $\alpha '$ or $\beta ' $. 
The uniformity is evident for the case of the Fibonacci lattice as
illustrated in Fig.~\ref{fig2}.
A proof of the uniformity for a generic
case will be given in the next section. Note that the ratio of the
self-similarity (i.e., the scaling factor $\tau$) of a binary QL is
restricted to a quadratic irrational of the form (\ref{unimodular3})
given in Appendix \ref{unimodular matrix}.~\cite{s1}

 Since the hyperscaling symmetry of the mother lattice is essential for
the self-similarity of the relevant binary QLs, we shall confine our
considerations to mother lattice with hyperscaling symmetries. Then,
every binary QL obtained by the projection method from such a 2D lattice
is self-similar, so that binary QLs appearing in the present paper are
all self-similar. Therefore, we may call a self-similar binary QL simply
as a binary QL.
 
 It is evident that $T^n$ with ${}^\forall n \in {\bf Z}$ transforms a
canonical set of basis vectors into another. We say that two canonical
sets are scaling-equivalent if one of them is transformed by $T^n$ with
${}^\exists n \in {\bf Z}$ into the other. Then, the relevant two binary
QLs are mutually similar, and we need not distinguish them in a
classification of QLs. The canonical windows of two QLs which are
scaling-equivalent are related to each other by $W'={\tau}^{-n}W$ with
${}^\exists n \in {\bf Z}$.
 
 Let ${\cal W}$ be the set of all the canonical windows. Then, it is a
discrete subset of the half line ${\bf R}^+$. It has the scaling
symmetry, ${\tau}^{\pm 1}{\cal W} = {\cal W}$, and is divided into
scaling-equivalent classes. Two binary QLs with windows belonging to
different scaling-equivalent classes are not mutually similar.

The parameter $\xi$ characterizing a binary QL satisfies $\xi > 1$ and
${\tilde \xi} > 0$, so that is a quasi-reduced quadratic irrational
(QRQI) as defined in Refs.\ \onlinecite{s2,s3} and discussed in Appendix
\ref{QRQI}. Conversely, the companion matrix of every QRQI is a
quasi-normal UFM, which determines a 2D lattice with a hyperscaling
symmetry and the relevant binary QL. Thus, there exists a bijection
(one-to-one correspondence) between any pair of the three sets: i) the
set of all the QRQIs, ii) the set of all the quasi-normal UFMs, and iii)
the set of all the binary QLs. In what follows, a binary QL specified by
a QRQI $\xi$ will be denoted as $Q\{\xi \}$. Since an inflation matrix
is a quasi-normal UFM, it determines uniquely a binary QL (exactly, an
LI class of binary QLs). Note that every quasi-normal UFM can be an
inflation matrix of a binary QL.

\section{Classification of Binary Self-Similar QLs into MLD Classes}
\label{sec_mld}

 If, between two QLs $Q$ and $Q'$, there exists a uniform local-rule for
 the transformation of $Q$ into $Q'$, $Q'$ is said to be {\it locally
 derivable} from $Q$; moreover, if $Q$ is also locally derivable from
 $Q'$, we say that $Q$ and $Q'$ are {\it mutually locally-derivable}
 (MLD).~\cite{comment4}
 
 According to a general theory of MLD relationship developed in Ref.\
 \onlinecite{cite2}, a necessary condition for two QLs to be MLD from
 each other is that they are derived by the projection method from a
 common mother lattice, where two mother lattices which differ in the
 scales of the physical space and the internal space are not
 distinguished as mentioned at the end of Sec.\ \ref{Binary
 QLs}.~\cite{comment5} Therefore, two binary QLs which are MLD but are
 not mutually similar are derived by two canonical sets, 
 $\{{\bf a},~{\bf b}\}$ and $\{{\bf a}',~{\bf b}'\}$, which are
 inequivalent with respect to the hyperscaling. Let $X$ be a unimodular
 matrix which combines two canonical sets:
 \begin{equation}
  ({\bf a}' \;\;~{\bf b}') = ({\bf a} \;\;~{\bf b})X. \label{modX}
 \end{equation}
Then, the argument in Appendix \ref{modular transformation} results in
that the ratio of the two intervals, $\xi' := b'/a'$, is related to the
ratio, $\xi := b/a$, by the modular transformation: $\xi' =
X^{-1}(\xi)$; the two ratios are modular equivalent. This is a necessary
condition for two binary QLs, $Q\{\xi\}$ and $Q\{\xi'\}$, to be MLD from
each other. The next problem is to show that this is also a sufficient
condition.

Using Eqs.\ (\ref{window}) and Eq.\ (\ref{modX}), we can calculate the
size of the canonical window $W'$ as $W'= |(t - u)\zeta - v + w|$, so
that the difference $W - W'$ with $W= 1 + \zeta$ belongs to the module
${\bf Z}[\zeta]$, where the relevant UFM $X$ is assumed to be given by
Eqs.\ (\ref{X}). Hence we can conclude by a general theory of the
MLD-relationship among different QLs that $Q\{\xi\}$ and $Q\{\xi'\}$ are
MLD.~\cite{NF} This completes a proof of the sufficiency
of the condition above. We will show in the next section directly the
MLD-relationship between $Q\{\xi\}$ and $Q\{\xi'\}$ by deriving an
explicit {\it substitution rules} (SR) combining the two QLs. At all
events, there exists a bijection between the set of all the MLD classes
of binary QLs and the set of all the modular equivalent classes (MECs)
of QRQIs. Therefore, we can specify an MLD class by the corresponding
MEC, $\langle\langle k_{0},~k_{1}, ~\cdots, ~k_{n-1} \rangle\rangle$, of
QRQIs, or, equivalently, by the corresponding MEC of the quasi-normal
UFMs, where $n$ is the rank of the class. Thus, there exist $N
\;(=k_{0}+k_{1}+ \cdots +k_{n-1})$ binary QLs in the relevant MLD class,
and the $N$ binary QLs are: $Q\{\xi^{(j)}_k\}$ with $j = 0, ~ 1, ~2, ~
\cdots, ~n -1$ and $k = 0, ~1, ~2, ~\cdots, ~k_{j} -1$, where the
$\xi^{(j)}_k$ are QRQIs defined in Appendix \ref{QRQI}. We shall call
$N$ the {\it order} of the MLD class. Other important parameters of the
MLD class is $m$ and $e = (-1)^n$, where $m$ is the common trace of the
relevant UFMs.

It is evident that each MLD class has its own mother lattice. That is,
there exists a bijection between the set of all the MLD classes of
binary QLs and the set of all the mother lattices with hyperscaling
symmetries.

\begin{figure}
  \begin{center}
  \centerline{\epsfig{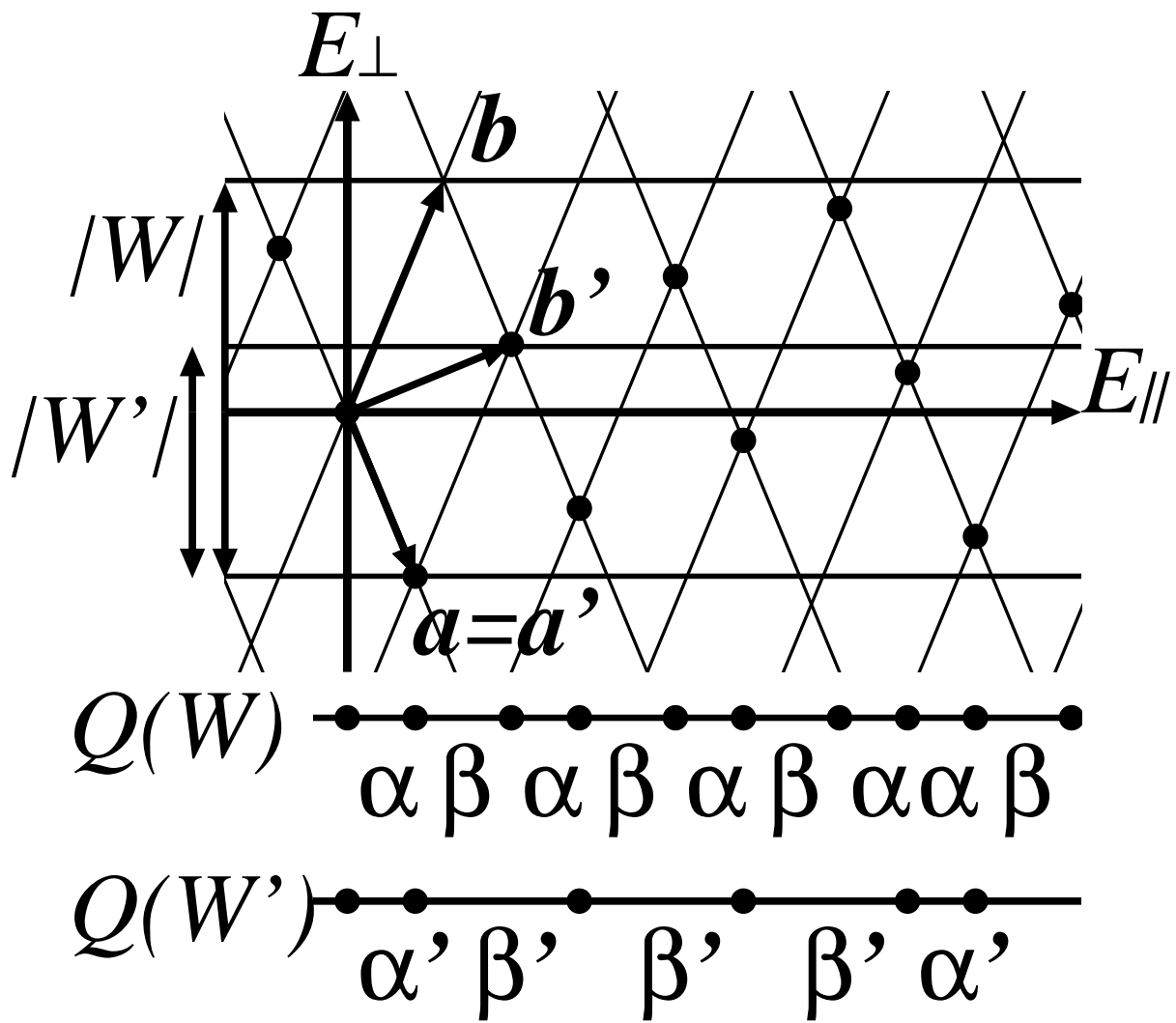}}
  \caption{A geometrical illustration for the SR (\ref{ruleS}). This
  case is different from the one illustrated in Fig.~\ref{fig2} in that
  the third basis vector ${\bf a}+{\bf b}$ is paired with ${\bf a}$ but
  not ${\bf b}$ to form a canonical set.}  \label{fig3}
   \end{center}
 \end{figure}

 The MLD-relationship between two QLs, $Q\{\xi\}$ and $Q\{\xi'\}$, in an
MLD class is particularly simple for the case where the modular
transformation, $\xi' = X^{-1}(\xi)$, coincides with either of the two
elementary modular transformations presented in Appendix \ref{continued
fractions}. The matrix $U$ given by Eq.\ (\ref{SUM}) is identical to the
matrix $M$ given by Eq.\ (\ref{M}), and the two canonical sets in this
case are related to each other by ${\bf a}' = {\bf b}$ and ${\bf b}' =
{\bf a}+{\bf b}$. Then, the SR relating the two sets of the intervals is
the same as the SR (\ref{inflation_fibonacci}) of the Fibonacci lattice:
 \begin{equation}
  \alpha'=\beta, \quad \beta'=\alpha\beta. \label{ruleU}
 \end{equation}
 This SR is closely related to the matrix $U$, so that we may call $U$
the substitution matrix of this SR. On the other hand, the two canonical
sets are related, in the case of the matrix $S$ given by Eq.\
(\ref{SUM}), to each other by ${\bf a}' = {\bf a}$ and ${\bf b}' = {\bf
a}+{\bf b}$. Then, we can show by a similar geometrical argument to the
one by which the SR above is derived that the SR for this case is given
by
 \begin{equation}
  \alpha'=\alpha, \quad \beta'=\beta\alpha, \label{ruleS}
 \end{equation}
 which is illustrated in Fig.~\ref{fig3}.
 The substitution matrix for this SR is $S$.

 A simple geometrical rule for the choice between the two SRs
 (\ref{ruleU}) and (\ref{ruleS}) is given as follows: the rule
 (\ref{ruleU}) or (\ref{ruleS}) should be chosen according as $\{{\bf
 b}, ~{\bf c}\}$ or $\{{\bf a}, ~ {\bf c}\}$ with ${\bf c} := {\bf
 a}+{\bf b}$ is a canonical set; ${\bf c}$ is the diagonal vector of the
 relevant parallelogram. This provides us with an elementary procedure
 which allows successive generation of canonical sets of basis
 vectors. This successive procedure can be regarded, alternatively, as a
 recursive procedure of generating a series of ``canonical'' basis
 vectors, ${\bf a}_i$, $i = 0, ~1, ~2, ~\cdots$, with the initial
 conditions, ${\bf a}_0 = {\bf a}$ and ${\bf a}_1 = {\bf b}$. We shall
 call it an additive algorithm. The inverse procedure of it is a
 subtractive algorithm: one and only one of $\{{\bf d}, ~{\bf a}\}$ and
 $\{{\bf d}, ~ {\bf b}\}$ with ${\bf d} := {\bf b}-{\bf a}$ is a
 canonical set, where ${\bf d}$ is the second diagonal vector of the
 relevant parallelogram. The additive algorithm (or the subtractive
 algorithm) may be called the diagonal (or anti-diagonal) algorithm. At
 any rate, we can generate a both-infinite set of canonical basis
 vectors $\Gamma := \{{\bf a}_i\,|\, i \in {\bf Z}\}$, and the $i$-th
 canonical set of basis vectors is given by $\{{\bf a}_k, ~ {\bf
 a}_i\}$, where $k$ is the maximum number under the condition that it is
 smaller than $i$ and the sign of the second component of ${\bf a}_k$ is
 opposite to that of ${\bf a}_i$. Hence there is a bijection between
 $\Gamma$ and the set of all the canonical sets of basis vectors, and we
 can identify them. The hyper-scaling symmetry of $\Lambda$ results in
 ${\bf a}_{i+N} = T{\bf a}_i$, so that $\Gamma$ is divided into $N$
 scaling-equivalent classes (series), which form an $N$-cycle, i.e., a 
 cyclically ordered set of $N$ elements.
 
 The properties of the trinity, $\{\Lambda, ~\Lambda_{\parallel}, ~
\Lambda_{\perp}\}$, is naturally succeeded by the trinity, $\{\Gamma, ~
\Gamma_{\parallel}, ~\Gamma_{\perp}\}$, where $\Gamma_{\parallel} :=
\{a_i\,|\, i \in {\bf Z}\}$ and $\Gamma_{\perp} := \{a^{\perp}_i\,|\, i
\in {\bf Z}\}$ are the projections of $\Gamma$ onto $E_{\parallel}$ and
$E_{\perp}$, respectively. In particular, $\Gamma_{\parallel}$ and
$\Gamma_{\perp}$ have the scaling symmetry, $a_{i+N} = \tau a_i$ and
$a^{\perp}_{i+N} = {\bar \tau} a^{\perp}_i$, and are divided into $N$
scaling-equivalent classes, which form $N$-cycles. The former of the
two is only composed of positive numbers but the latter is not. If the
signs of the members are ignored, $\Gamma_{\perp}$ coincides with ${\cal
W}$ because each member of $\Gamma$ is the ``anti-diagonal'' of the
parallelogram associated with a canonical set of basis vectors. If
$\{{\bf a}_k, ~{\bf a}_i\}$ is a canonical set above, we obtain $\xi_i =
a_i/a_k$ and $\zeta_i = - a^{\perp}_k/a^{\perp}_i$, which are periodic:
$\xi_{i+N} = \xi_i$ and $\zeta_{i+N} = \zeta_i$. A 2D matrix is defined
with the canonical set by $A_i := ({\bf a}_k ~~{\bf a}_i)$. It satisfies
$A_{i+1} = A_iS$ if $a^{\perp}_i$ and $a^{\perp}_{i+1}$ have a common
sign but $A_{i+1} = A_iU$ otherwise. It satisfies, furthermore, $TA_i =
A_{i+N} = A_iM$, where $M$ is the companion UFM of $\xi$.

We may write $A_i = ({\bf a} \;\;~{\bf b})X_i$, where
 \begin{equation}
 X_i := \left( \begin{array}{cc} t_i & u_i \\ v_i & w_i
 \end{array} \right) \label{Mi}
 \end{equation}
is a unimodular matrix. It is a quasi-normal UFM if $i \ge 2$ but the
reciprocal of a quasi-normal UFM if $i \le 0$, while $X_1 = I$. It
satisfies $X_{i+N} = X_iM$. Since $A_i = ({\bf a}_k ~~{\bf a}_i)$, we
obtain ${\bf a}_i = u_i{\bf a}+w_i{\bf b}$, $a_i = u_i+w_i\xi \in {\bf
Z}[\xi]$, and $a^{\perp}_i = -u_i\zeta + w_i \in {\bf Z}[\zeta]$. The
series of rationals, $w_i/u_i$ (or $-u_{-i}/w_{-i}$), $i = 1, ~2, ~
\cdots$, is just the series of best approximants to $\zeta$ (or
$\xi$). The procedure to obtain approximants to an irrational on the
basis of the geometry of a lattice has been established in the classical
number theory.~\cite{Takagi}
 
 It is sometimes convenient to change a single suffix numbering, $x_i$,
into a double suffix numbering, $x_n^{(j)}$, with $i = j + nN$, where $j
= 1, ~2, ~\cdots, ~N$ and $n \in {\bf Z}$. Then a relation combining
$x_i$ and $x_{i+N}$ turns to that combining $x_n^{(j)}$ and
$x_{n+1}^{(j)}$. Using the equality $X_n^{(j)} = X_0^{(j)}M^n$ together
with the Cayley-Hamilton theorem, $M^2 - mM + eI = 0$, we can conclude
that the quantities ${\bf a}_n^{(j)}$, $A_n^{(j)}$, $X_n^{(j)}$,
$a_n^{(j)}$, and $a_n^{\perp, (j)}$ satisfy two-term recursion relations
which are isomorphic to $x_{n+1}^{(j)} = mx_{n}^{(j)} - ex_{n-1}^{(j)}$;
the solution of the recursion relation is written as $x_n = F_nx_1 -
eF_{n-1}x_0$ with $F_n$ being the generalized Fibonacci numbers
associated with the quadratic irrationals $\tau$.~\cite{comment3}
Remember, however, that the double suffix numbering must not apply to
$\xi$ and $\zeta$.

The $N$ QRQIs in the MEC introduced above form an $N$-cycle of modular
transformations, and the relation between successive two members of the
cycle is given by
 \begin{equation}
 \xi^{(j)}_k = S(\xi^{(j)}_{k-1}) \label{S-modular}
 \end{equation}
  if $k > 0$ but
 \begin{equation}
  \xi^{(j)}_0 = U(\xi^{(j-1)}_{k}) \label{U-modular}
 \end{equation}
 with $k = k_{j-1} -1$. It follows that successive two members of the
cycle of $N$ binary QLs are related by the SR (\ref{ruleU}) or
(\ref{ruleS}). Moreover, if two members of the $N$-cycle are chosen
arbitrarily, each of the two is derived from the other by a successive
operation of the two types of the elementary SRs. That is, the two QLs
are combined by composite SRs.

It is appropriate at this point to define exactly the SR which combines
two QLs. We shall distinguish for a while the two types of intervals
$\alpha '$ or $\beta '$ from two types of sequences $\sigma_\alpha$ or
$\sigma_\beta$ composed of the two types of intervals $\alpha$ or
$\beta$. An SR is represented as $\alpha ' = \sigma_\alpha$, $\beta ' =
\sigma_\beta$. Let us denote a QL composed of the two types of intervals
$\alpha$ or $\beta$ as $Q[\alpha, \;\beta]$. If $Q[\alpha, \;\beta]$ is
represented as a sequence of two types of sequences $\sigma_\alpha$ or
$\sigma_\beta$, there exists a new QL $Q'[\alpha ', \;\beta ']$, so that
$Q[\alpha, \;\beta] = Q'[\sigma_\alpha, \;\sigma_\beta]$. If the SR is
used to derive $Q'[\alpha ', \;\beta ']$ from $Q[\alpha, \;\beta]$, we
may call it a passive SR. Contrary to it, if it is used to derive the
latter from the former, we may call it an active SR. The passive SR and
the active one are the inverse procedure of each other.

The substitution matrix of any SR is naturally a quasi-normal UFM. If
the QL $Q\{\xi\}$ is transformed by a passive SR into $Q\{\xi'\}$, the
relevant two quadratic irrationals is related by the modular
transformation as $\xi = X(\xi')$ with $X$ being the relevant
substitution matrix.  If an SR with the substitution matrix $X$ is
operated first, and the second one with $X'$ is operated subsequently,
the substitution matrix of the resulting composite SR is given by $XX'$
in the passive case but $X'X$ in the active case. Remember that, for the
passive case, the composition of two SRs is anti-isomorphic to the
product of the two relevant substitution matrices. In the present paper,
we will use SRs in the passive meaning unless stated otherwise.

\begin{figure}
  \begin{center}
  \centerline{\epsfig{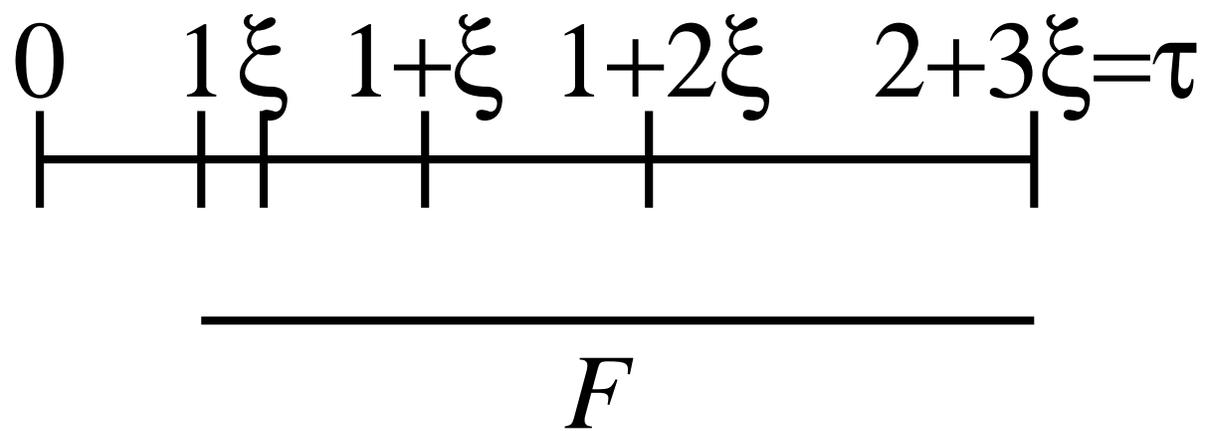}}
  \caption{The ``finger print'' of the MLD class $\langle\langle
  1,~2,~1 \rangle\rangle$.}  \label{fig4}
   \end{center}
 \end{figure}

 Each of the $N$ scaling-equivalent classes in ${\cal W}$ has one and
only one member in the fundamental interval, $F :=\; ]1, ~\tau]$. Let
${\cal W}_0 := {\cal W}\cap{\bar F}$ with ${\bar F} :=[1, ~\tau]$. Then,
it includes $N + 1$ windows. The pattern of the distribution of the
windows in ${\cal W}_0$ characterizes completely the relevant MLD class,
so that ${\cal W}_0$ is like a finger print of the MLD class. For
example, if $\xi = \langle 1,~2,~1 \rangle = (1+\sqrt{10}\,)/3$, we
obtain ${\cal W}_0 = \{1, ~\xi, ~1 + \xi, ~1 + 2\xi, ~2 + 3\xi\}$, which
is shown in Fig.~\ref{fig4}.
The last member of this ``finger print'' is
equal to $\tau = 3+\sqrt{10}\,$, which is scaling-equivalent to the
first member.

 In general, ${\bar F}$ is divided into $N$ subintervals by ${\cal
W}_0$. Every subinterval is limited by two canonical windows, and the
two relevant binary QLs have one common interval. If a window is chosen
to be an internal point of the subinterval, the resulting QL is a
ternary QL composed of the three types of intervals associated with the
two binary QLs.~\cite{NF}That is, the ternary QL is a hybrid of the two
binary QLs.
 
 The canonical windows are vertical widths of the fundamental
parallelograms, and we may call their horizontal widths canonical
interval (lengths) by a reason to be understood shortly. Let us denote
by ${\cal V}$ the set of all the canonical intervals. Then it is
identical to $\Gamma_{\parallel}$. The sizes of the two intervals
included in a binary QL derived from the relevant mother lattice are
given by consecutive two members of ${\cal V}$. A new ``finger print''
${\cal V}_0$ can be defined with respect to the scaling-invariant set
${\cal V}$ in a similar way as the one for ${\cal W}_0$. It is
remarkable that the two types of ``finger prints'' coincide for the MLD
classes like $\langle\langle 1,~2,~1 \rangle\rangle$. The reason will be
revealed in a later section.

\section{Composite Substitution Rules and Inflation Rules} \label{sec:composite}

 It is sometimes convenient to employ composite SRs associated with the
two types of UFMs which are given as blocked forms, $L = S^{l}$ and $K =
US^{k-1}$ with $k$ and $l$ being natural numbers. The SR for the former
case is $l$ fold repetition of the rule (\ref{ruleS}), so that we obtain
\begin{equation}
 \alpha'=\alpha, \quad \beta'=\beta\alpha^{l}, \label{ruleS^{l}}
\end{equation}
where $\alpha^{l}$ stands for a concatenation of $l$ $\alpha$s. This SR
is a direct consequence of the equations, ${\bf a}' = {\bf a}$ and ${\bf
b}' = l{\bf a}+{\bf b}$. The explicit form of the substitution matrix of
this SR is given by Eq.\ (\ref{L}) in Appendix \ref{Frobenius}. On the
other hand, the SR for the substitution matrix $K = US^{k-1}$ is a
composition of the rule (\ref{ruleS^{l}}) with $l = k - 1$ and the rule
(\ref{ruleU}), so that we obtain
\begin{equation}
 \alpha'=\beta, \quad \beta'=\alpha\beta^{k}. \label{ruleUS^{k}}
\end{equation}
This SR is a direct consequence of the equations, ${\bf a}' = {\bf b}$
and ${\bf b}' = {\bf a}+k{\bf b}$. The two block SRs just obtained have
simple geometrical meanings, which can also be understood easily. Since
a quasi-normal UFM is represented as Eq.\ (\ref{M'}), every composite SR
can be represented as a composition of a number of block SRs of the
types (\ref{ruleS^{l}}) and (\ref{ruleUS^{k}}).

 A composition of two SRs as well as a product of the corresponding two
substitution matrices is incommutable. For example, the SR with the
substitution matrix $SU$ is given by
\begin{equation}
 \alpha'=\beta\alpha, \quad \beta'=\alpha\beta\alpha, \label{SU}
\end{equation}
which is markedly different from the SR with the substitution matrix
$US$; this SR is given by Eq.\ (\ref{ruleUS^{k}}) with $k = 2$.

 Let $Q$ and $Q'$ be two QLs which are combined by the SR
(\ref{SU}). Then, they are combined also by another SR,
$\alpha'=\alpha\beta$, $\beta'=\alpha^2\beta$, which is obtained from
the former by moving the interval $\alpha$ at the right ends of the two
relevant sequences of the intervals to their left ends. Therefore, these
two SRs are equivalent. By similar procedures, we can obtain another two
equivalent SRs: $\alpha'=\beta\alpha$, $\beta'=\beta\alpha^2$ and
$\alpha'=\alpha\beta$, $\beta'=\alpha\beta\alpha$. Conversely, by
applying repeatedly the inverse procedure to the last SR, we can
retrieve other three SRs. These proper and inverse procedures are called
cyclic shifts.~\cite{NF} They yield a linearly ordered finite set of
equivalent SRs, and the set is retrieved from any member of the set. The
SRs in the set have a common substitution matrix. We will not
distinguish equivalent but apparently different SRs. Note that the full
set of equivalent SRs has a sort of mirror symmetry; an SR in the
ordered set and the one at the mirror site are mutually mirror symmetric
because each sequence of the former is the mirror image of the relevant
sequence of the latter.

 An MLD class of binary QLs has the structure of an $N$-cycle as
mentioned in the preceding section. Therefore, if the two types of
elementary SRs are applied $N$ times in an appropriate order to an
arbitrarily chosen member of the $N$-cycle, the chosen member is
retrieved. In other words, every binary QL is transformed into itself by
a composite SR. Therefore, it is self-similar, and the composite SR is
nothing but the IR. Since the inflation matrix of a QL is written as
Eq.\ (\ref{M'}), the relevant IR is obtained by a composition of the two
types of the block SRs (\ref{ruleS^{l}}) and (\ref{ruleUS^{k}}).

\section{Examples} \label{sec_examples}
\subsection{MLD classes of rank one} \label{Rank One}

 MLD classes of rank one form a series specified by natural numbers, and
most of important 1D QLs including the Fibonacci lattice belong to some
MLD classes of this series. The MLD class is specified by
$\langle\langle m \rangle\rangle$ in our notation, for which $N = m$ and
$e = -1$. It has only one RQI as given by
  \begin{equation}
   \tau=\frac{1}{2}\left(m+ \sqrt{m^{2}+4} \,\right), \label{TAU}
 \end{equation}
 i.e., the precious mean, which presents the scale of the
self-similarity of the binary QLs in the class. There exist $m$ QRQIs,
$\xi^{(0)}_{k}= \tau - k$ with $k=0, ~1, ~\cdots, ~m-1$.

 The inflation matrix for $Q\{\xi^{(0)}_{k}\}$ is written with Eq.\
 (\ref{M'}) as
 \begin{equation}
  M_{k}=S^{k}US^{m-k-1}= \left(
   \begin{array}{cc}
    k & k(m-k)+1 \\ 1 & m-k
   \end{array} \label{inflmtrxMk} 
   \right).
 \end{equation}
 The IR for $Q\{\xi^{(0)}_{k}\}$ is given as a composition of two block
 SRs corresponding to $US^{m-k-1}$ and $S^{k}$. Using Eqs.\
 (\ref{ruleS^{l}}) and (\ref{ruleUS^{k}}), we obtain
 \begin{equation}
  \alpha'=\beta\alpha^{k}, \quad \beta'=\alpha(\beta\alpha^{k})^{m-k},
  \label{rule-m}
 \end{equation}
 which is equivalent to the IR given in Ref.\ \onlinecite{NF} or given
by equation (20) in Ref.\ \onlinecite{cite7}.

There exists only one MLD class for $N = 1$ or $N = 2$ but there exist two
or more for $N \ge 3$. The MLD class $\langle\langle 1 \rangle\rangle$ is
only composed of the Fibonacci lattice, while $\langle\langle 2
\rangle\rangle$ is composed of two MLD classes whose ratio of
self-similarity is the silver mean, $\tau = 1 + \sqrt{2}\,$. The IRs of the
two binary QLs in $\langle\langle 2 \rangle\rangle$ are given by the above
equations with $k = 0$ and $k = 1$, and the relevant quadratic irrationals
are $\xi^{(0)}_{0} = 1 +\sqrt{2}\,$ and $\xi^{(0)}_{1} = \sqrt{2}\,$,
respectively. The two QLs are transformed to each other by the elementary
SRs, (\ref{ruleU}) and (\ref{ruleS}).

 The ratio of self-similarity, $\tau$, of a QL must be generally a
 positive root of the quadratic equation $z^{2}-m z +e=0$ with $m$ being
 a natural number and $e = \pm1$. There exists at least one MLD class
 with given $m$ and $e$. A representative of such MLD classes is
 $\langle\langle m \rangle\rangle$ if $e = -1$. There exists only one
 MLD class with given $m$ and $e$ if $m \le 5$ and $e = -1$. The two MLD
 classes, $\langle\langle 6 \rangle\rangle$ and $\langle\langle 1,~2,~1
 \rangle\rangle$, are commonly characterized by $m = 6$ and $e = -1$
 (namely, $\tau=3+\sqrt{10}\,$).

 A remarkable feature of binary QLs of rank one is that the relevant
parameters satisfy the condition, ${\bf Z}[\xi] = {\bf Z}[\tau]$, which
results in that the relevant QLs have simple IRs.

\subsection{MLD classes of rank two}
 MLD classes of rank two form a 2D ``series'', $\langle\langle k_0,~k_1
\rangle\rangle$, each of which is specified by a pair of natural
numbers, and we obtain $e = +1$. We can restrict our considerations to
the case $ k_0 > k_1$. An important subseries is formed of
$\langle\langle m-2,~1 \rangle\rangle$ with $m \ge 4$, for which $N = m -
1$. This subseries has been investigated by several authors (see, for
example, Ref.\ \onlinecite{NF}).

 The companion matrix $M_0 = K_{1}K_{0}$ of the RQI $\theta_{0} :=
\langle m-2,~1 \rangle$ is given by
 \begin{equation}
  M_{0}= \left(
   \begin{array}{cc}
    1 & m-2 \\ 1 & m-1
   \end{array}
   \right), \label{inflmtrxM0}
 \end{equation}
 whose Frobenius eigenvalue is
  \begin{equation}
  \tau=\frac{1}{2}\left(m+ \sqrt{m^{2}-4} \,\right),
 \end{equation}
   i.e., the ``anti-precious mean''.~\cite{cite7} It presents the scale
of the self-similarity of the binary QLs in the relevant class, and
$\theta_{0}$ is written as $\theta_{0} = \tau - 1$. The relevant MEC of
QRQIs has $m-1$ members, $\xi^{(0)}_{k}:=\tau - k$ with $k=1, 2, ~
\cdots, ~m-2$ and $\xi^{(1)}_{0}=\langle 1, m-2 \rangle = (\tau -
1)/(m-2)$, where the parameter $k$ is shifted for a convenience from
that following the convention in Sec.\ \ref{sec_mld}. The inflation
matrices for the QLs, $Q\{\xi^{(0)}_{k}\}$ and $Q\{\xi^{(1)}_{0}\}$, are
$S^{k-1}U^{2}S^{m-k-2}$ and $US^{m-3}U \;(= {}^{\rm t}\!M_0)$,
respectively. The IR for $Q\{\xi^{(0)}_{k}\}$ is shown to be equivalent
to
 \begin{equation}
   \alpha'=\alpha^{k}\beta, \quad
  \beta'=\alpha^{k-1}\beta(\alpha^{k}\beta)^{m-k-1}, \label{rule-xi0k}
 \end{equation}
 which is also equivalent to the IR given in Ref.\ \onlinecite{NF} or
given by equation (20) in Ref.\ \onlinecite{cite7}. On the other hand,
the IR for $Q\{\xi^{(1)}_{0}\}$ is shown to be equivalent to
 \begin{equation}
  \alpha'=\beta^{m-2}\alpha, \quad
  \beta'=\beta^{m-1}\alpha. \label{rule-m-1}
 \end{equation}
 If the notations for the two types of intervals are exchanged in this
IR, the result coincides with the IR (\ref{rule-xi0k}) but with $k =
m-1$. Therefore, the formula (\ref{rule-m-1}) is absorbed in
(\ref{rule-xi0k}).~\cite{NF} Note, however, that the second interval is
shorter than the first one after the exchange of the notations. The
explicit form of the inflation matrices are:
 \begin{equation}
  M_{k}=S^{k-1}U^{2}S^{m-k-2}= \left(
   \begin{array}{cc}
    k & k(m-k)-1 \\ 1 & m-k
   \end{array} \label{M-m-1} 
   \right).
 \end{equation}

 The simplest MLD class of the subseries $\langle\langle m-2,~1
\rangle\rangle$ is given by setting $m =4$. It includes three binary QLs
whose ratio of self-similarity is $\tau=2+\sqrt{3}$.

 The inflation matrices of binary QLs of rank two have the property,
${\bf Z}[\xi] = {\bf Z}[\tau]$ or ${\bf Z}[\xi^{-1}] = {\bf Z}[\tau]$,
only if they belong to the subseries above but this property is not
possessed by other binary QLs of rank two nor those of higher ranks.

 A representative of MLD classes with given $m$ but with $e = +1$ is
$\langle\langle m-2,~1 \rangle\rangle$. There exist no other MLD classes
if $m \le 7$. The two MLD classes, $\langle\langle 7,~1 \rangle\rangle$
and $\langle\langle 3,~2 \rangle\rangle$, are commonly characterized by
$m = 8$ and $e = +1$ (namely, $\tau=4+\sqrt{15}\,$). In fact, the second
MLD class is the simplest among rank two MLD classes which do not belong
to the subseries above. It has five binary QLs corresponding to five
QRQIs:
 \begin{eqnarray*}
  \xi^{(0)}_{0} &=& \frac{3+\sqrt{15}}{2}, \quad \xi^{(0)}_{1} =
  \frac{1+\sqrt{15}}{2}, \quad \xi^{(0)}_{2} = \frac{-1+\sqrt{15}}{2},
  \\ \xi^{(1)}_{0} &=& 1+\frac{\sqrt{15}}{3}, \quad \xi^{(1)}_{1} =
  \frac{\sqrt{15}}{3}.
 \end{eqnarray*}
 
 The MLD class $\langle\langle 3,~2 \rangle\rangle$ is considered to be
the simplest member of the series of rank two MLD classes,
$\langle\langle k,~2 \rangle\rangle$ with $k \ge 3$. The relevant
parameters are: $m = 2k + 2$, $e = +1$, and $N = k + 2$. An important QL
in the MLD class $\langle\langle k,~2 \rangle\rangle$ is
$Q\{\xi^{(1)}_{1}\}$, whose inflation matrix is $S(US^{k-1}US)S^{-1} =
SUS^{k-1}U$. Explicitly, the inflation matrix coincides with a
quasi-normal UFM defined by
 \begin{equation}
  M_k^+ := \left(
   \begin{array}{cc}
    k+1 & k+2 \\ k & k+1
   \end{array}
   \right), \label{MYa1}
 \end{equation}
 so that $\theta_{0} = (\tau - k - 1)/k$. Since $M$ is factorized into
three blocks, $(S)(US^{k-1})(U)$, the relevant IR is obtained as a
composition of the three relevant SRs:
 \begin{equation}
  \alpha'=\alpha(\beta\alpha)^{k}, \quad
\beta'=\alpha(\beta\alpha)^{k+1}. \label{IRYa1}
 \end{equation}
 A series of binary quasiperiodic sequences with this type of IRs has
 been investigated in Ref.\ \onlinecite{Ya93}. Note, however, that the
 sequences in Ref.\ \onlinecite{Ya93} are not geometrical objects but
 those of letters.
 
\subsection{MLD classes of rank three}

 An important series of QLs formed of rank three MLD classes is given by
$\langle\langle 1,~k,~1 \rangle\rangle$ with $k \ge 2$. The relevant
parameters are: $m = 2k + 2$, $e = -1$, and $N = k + 2$. The companion
matrix $M_0 = UKU= U^2S^{k-1}U$ of the RQI $\theta_{0} := \langle
1,~k,~1 \rangle$ coincides with the quasi-normal UFM defined by
 \begin{equation}
  M_k^- := \left(
   \begin{array}{cc}
    k & k+1 \\ k+1 & k+2
   \end{array}
   \right), \label{MYa2}
 \end{equation}
 so that $\theta_{0} = (\tau - k)/(k + 1)$. The IR for
$Q\{\theta_{0}\}$, for example, is given by
 \begin{equation}
  \alpha'=\beta(\alpha\beta)^{k}, \quad \beta'=\beta(\alpha\beta)^{k+1}.
\label{IRYa2}
 \end{equation}
 A series of binary quasiperiodic sequences with this type of IRs has
 been also investigated in Ref.\ \onlinecite{Ya93}. The simplest MLD
 class in the series is $\langle\langle 1,~2,~1 \rangle\rangle$, which
 has appeared at the end of the subsection \ref{Rank One}. The inflation
 matrix given by Eq.\ (\ref{inflmtrx2334}) is a special case of the
 above with $k = 2$.

\section{Symmetric Substitution Rules and Symmetric Inflation
Rules}\label{sec:symmetric}
 
 The composite SR (\ref{ruleUS^{k}}) with $k = 2$ is equivalent to
 \begin{equation}
  \alpha'=\beta, \quad \beta'=\beta\alpha\beta, \label{US}
 \end{equation}
 which is explicitly symmetric, and we may say that the SR
(\ref{ruleUS^{k}}) with $k = 2$ is symmetric. More generally, the SR
(\ref{ruleUS^{k}}) is symmetric if $k$ is even but not if $k$ is odd. A
similar argument applies to the SR (\ref{ruleS^{l}}). Explicitly, we may
write
 \begin{equation}
  \alpha'=\alpha, \quad \beta'=\alpha^{l'}\beta\alpha^{l'} \label{S2l}
 \end{equation} 
 if $l = 2l'$ with $l'$ being a natural number. In particular, if $l =
 2$, it reduces to
 \begin{equation}
  \alpha'=\alpha, \quad \beta'=\alpha\beta\alpha. \label{S2}
 \end{equation} 
On the other hand, the two elementary SRs (\ref{ruleU}) and
(\ref{ruleS}) are asymmetric. Also the SR (\ref{SU}) is asymmetric,
while the two IRs, (\ref{IRYa1}) and (\ref{IRYa2}), are symmetric
irrespective of the value of $k$. We should remark that the two SRs,
(\ref{US}) and (\ref{S2}), have simple geometrical interpretations in
the projection method.
 
 We shall consider here the condition for an SR to be symmetric.  A
sequence composed of two types of intervals, $\alpha$ and $\beta$,
cannot be symmetric if both the number of $\alpha$ in it and that of
$\beta$ are odd.  Thus, the parities of these numbers are important. It
can be readily shown that a necessary condition for an SR to be
symmetric is that its substitution matrix is a quasi-permutation-matrix
(QPM) as defined in Appendix \ref{quasi-normal}. This condition is shown
to be a sufficient condition as well.~\cite{NF} This is consistent with
above observations. For example, the inflation matrix (\ref{inflmtrxMk})
is a QPM if and only if both $m$ and $k$ are even, so that the IR
(\ref{rule-m}) is symmetric only when this condition is
satisfied. Explicitly, the symmetric IR is equivalent to $\alpha' =
\alpha^{k'}\beta\alpha^{k'}$ and $\beta' =
(\alpha')^{m'-k'}\alpha(\alpha')^{m'-k'}$ with $m' := m/2$ and $k' :=
k/2$. A similar argument applies to the IR (\ref{rule-xi0k}).

It is shown in Appendix \ref{quasi-normal} that $M^2$ with $M$ being an
inflation matrix is always a QPM if $m$ is even. Since $M^2$ is
associated with the double IR, the double IR is equivalent to a
symmetric IR. Similarly, the triple IR is equivalent to a symmetric IR
if $m$ is odd. Note that the scaling ratio of the double or triple IR is
equal to $\tau^2$ or $\tau^3$, respectively. For example, if the
symmetric IR given by Eq.\ (\ref{SU}) is doubled, we obtain the
symmetric IR: 
 \begin{equation}
  \alpha' = \alpha(\beta\alpha)^2$, \quad $\beta' =
\alpha(\beta\alpha)^3, \label{double}
 \end{equation} 
whose ratio of inflation-symmetry is $(1 + \sqrt{2}\,)^2$. On the other
hand, if the IR of the Fibonacci lattice is tripled, we obtain the
symmetric IR: $\alpha' = \beta\alpha\beta$ and $\beta' =
\beta(\alpha\beta)^2$, whose ratio of inflation-symmetry is $\tau_{\rm
G}^3$.

 The central interval in the sequence $\alpha'$ of the symmetric IR
(\ref{IRYa2}) is $\alpha$ and that in $\beta'$ is $\beta$ if $k$ is odd
but these correspondences are reversed if $k$ is even. The relevant
substitution matrix is congruent in modulo 2 with $I$ for the former
case but with $J := \left({0 \atop 1}\; {1 \atop 0}\right)$ for the
latter. These observations are general properties of symmetric SRs and
symmetric IRs.
 
 If one of two QLs is derived from the other by a symmetric SR, and,
moreover, the converse is also true, we may say that they are
symmetrically MLD (s-MLD). Since the s-MLD is stronger
than the conventional MLD, an MLD class can be divided into two or more
s-MLD classes. For example, the MLD class $\langle\langle m
\rangle\rangle$ with an even $m$ necessarily bifurcates because it
includes two types of QLs with different scalings with respect to
symmetric IRs. Note that a necessary condition for the two QLs to be
s-MLD is that the relevant two windows is
concentric.~\cite{Ni89a}
 
 A symmetric SR is called elementary if it cannot be represented as a
composition of two symmetric SRs. Such an SR is associated with an
elementary QPM defined in Appendix \ref{quasi-normal}. In particular,
the SRs (\ref{US}) and (\ref{S2}) are elementary symmetric SRs. Each
s-MLD class is a cycle, and its two successive members are combined by
an elementary symmetric SR. There are two important series of elementary
QPMs, namely, (\ref{MYa1}) and (\ref{MYa2}) with $k \ge 0$. This is
confirmed by the expressions $M_k^+ = SUS^{k-1}U$ and $M_k^- =
U^2S^{k-1}U$ for $k \ge 1$ but $M_0^+ = S^2$ and $M_0^- = US$. The
corresponding symmetric SRs are identical to the IRs, (\ref{IRYa1}) and
(\ref{IRYa2}), respectively. That is, QLs belonging to the two series of
QLs with inflation matrices, $M_k^+$ with $k \ge 1$ and $M_k^-$ with $k
\ge 0$, have the property that their inflation matrices are
elementary. The two series are the main subjects in Ref.\
\onlinecite{Ya93}.

 From here to the end of this section, we use SRs in the active meaning
for a convenience; the directions of the arrows to appear should be
reversed in the case of the passive meaning. A simplest example is again
the case of a rank one MLD class, $\langle\langle m \rangle\rangle$. For
brevity we set $Q_k := Q\{\xi^{(0)}_{k}\}$, whose inflation matrix is
denoted as $M_k$. If $m$ is even, the inflation matrix for $Q_0$ is
factorized into elementary QPMs as $M_{0}\;(=US^{m-1})=
(M_0^-)(M_0^+)^{m/2-1}$, which yields the following cycle of symmetric
SRs:
 \begin{equation}
    Q_0 \rightarrow Q_2 \rightarrow \cdots \rightarrow Q_{m-2}
   \Rightarrow Q_0, \label{cycle0}
 \end{equation}
 where the single arrow stands for the SR (\ref{S2}) but the double
arrow the SR (\ref{US}). The above cycle only includes the QLs $Q_k$
with even $k$. For $Q_k$ with an odd $k$, the double IR must be
employed. The relevant inflation matrix for $Q_1$ is written as
$(M_{1})^2=SUS^{m-1}US^{m-2}= (M_{m}^+)(M_0^+)^{m/2-1}$, so that we
obtain the following cycle of odd $k$ QLs:
 \begin{equation}
    Q_1 \rightarrow Q_3 \rightarrow \cdots \rightarrow Q_{m-1}
   \Rightarrow Q_1, \label{cycle1}
 \end{equation}
 where the double arrow stands for the symmetric SR which is given by
(\ref{IRYa1}) with $k = m$. Thus, the MLD class $\langle\langle m
\rangle\rangle$ with an even $m$ is divided evenly into the even s-MLD
class and the odd s-MLD class. Each of the two s-MLD classes for the
case of $m = 2$ is composed of a single QL, $Q_0$ or $Q_1$. The
symmetric IR for $Q_0$ is given by (\ref{US}), while that for $Q_1$ by
(\ref{double}).
 
 We shall proceed to the case of the MLD class $\langle\langle m
\rangle\rangle$ with an odd $m$. Then triple IRs are in order. The
relevant inflation matrix for the case of $Q_0$ is factorized as
 \begin{eqnarray}
    (M_{0})^3 &=& (US^{m-1})^{3} \nonumber \\ &=&
     (M_0^-)(M_0^+)^{(m-3)/2}(M_{m}^+)(M_0^+)^{(m-1)/2},
     \label{cycleodd}
 \end{eqnarray}
which reduces to Eq.\ (\ref{cycletriple}) if $m = 5$. Since the right
hand side has $m$ factors, the $m$ QLs, $Q_k$ with $k=0, ~1, ~ \cdots, ~
m-1$, are cyclically related by elementary symmetric SRs:
  \begin{eqnarray}
   &&Q_0 \rightarrow Q_2 \rightarrow \cdots \rightarrow Q_{m-1}
    \Rightarrow Q_1 \rightarrow Q_3 \rightarrow \cdots \nonumber \\
    &\rightarrow& Q_{m-2} \Rightarrow Q_0. \label{cycle3}
   \end{eqnarray}
That is, this MLD class is simultaneously an s-MLD class.
 
 The above results for the case of a rank one MLD class is readily
extended to the case of the subseries $\langle\langle m-2,~1
\rangle\rangle$ of rank two MLD classes. We only present here the
results of factorizations of the relevant inflation matrices. The case
with an even $m$ has two s-MLD classes, and the relevant factorization
for the even s-MLD class is $M_{2} = (M_1^+)(M_0^+)^{m/2-2}$ but the one
for the odd s-MLD class is $(M_{1})^2 =
(M_{m-2}^-)(M_0^-)(M_0^+)^{m/2-2}$ The relevant factorization for an odd
$m$ is $(M_{1})^{3}=
(M_{m-2}^-)(M_0^-)(M_0^+)^{(m-5)/2}(M_1^+)(M_0^+)^{(m-3)/2}$. Note that
$Q_k$ with $k=0$ is forbidden in the present case, so that the even
s-MLD class with even $m$ is composed of $m/2 - 1$ QLs. The situations
are more complicated for other MLD classes of rank two or of higher
ranks.
 
 A quick derivation of a symmetric IR is possible if we use blocked SRs.
The SR (\ref{S2l}) is associated with the block $(S^2)^{l'}$. Another
two SRs associated with the two important blocks $(M_k^+)(M_0^+)^{p}$
and $(M_k^-)(M_0^+)^{p}$ are written with ${\hat \beta} :=
\alpha^{p}\beta\alpha^{p}$ as
 \begin{equation}
  \alpha'={\hat \beta}(\alpha{\hat \beta})^{k}, \quad \beta'={\hat
\beta}(\alpha{\hat \beta})^{k+1}, \label{sblock1}
 \end{equation}
 \begin{equation}
  \alpha'=\alpha({\hat \beta}\alpha)^{k}, \quad \beta'=\alpha({\hat
  \beta}\alpha)^{k+1}. \label{sblock12}
 \end{equation}
 
 Although the SR (\ref{SU}) is asymmetric, it is equivalent to the
symmetric but fractional SR:
 \begin{equation}
  \alpha'=(\frac{1}{2}\beta)\alpha(\frac{1}{2}\beta), \quad
\beta'=(\frac{1}{2}\beta)\alpha^2(\frac{1}{2}\beta). \label{fractional}
 \end{equation} 
 It can be shown generally that every asymmetric SR is equivalent to a
symmetric but fractional SR.~\cite{NF} The same is true for asymmetric
IRs.  Thus, an MLD class is simultaneously an s-MLD class if fractional
SRs are allowed.

\section{Miscellaneous Subjects}\label{sec:miscellanious}
 
\subsection{Periodic approximants}\label{approximant}

 Let $\alpha'$ and $\beta'$ be two types of sequences of intervals
$\alpha$ and $\beta$ of a QL, $Q$, and assume that the pair of the
sequences forms an SR. Then, the periodic structure $(\alpha')^\infty$
(or $(\beta')^\infty$) obtained by an infinite concatenation of
$\alpha'$ (or $\beta'$) is called a periodic approximant to
$Q$.~\cite{EH85} It is evident that the period of a periodic approximant
is equal to a canonical interval, and different periodic approximants
have different periods. Therefore, we shall denote by $Q_n^{(j)}$ ($n
\ge 0$) the periodic approximant corresponding to the canonical interval
$a_n^{(j)}$. The periodic approximant $Q_n^{(j)}$ is changed by an
active IR to $Q_{n+1}^{(j)}$. Hence all the periodic approximants are
derived by the successive application of the IR to the prototypes,
$Q_0^{(j)}$ with $j = 0, ~1, ~2, ~\cdots, ~N-1$.~\cite{N91} The first
three prototypes are $\alpha$, $\beta$, and $\alpha\beta$, whose periods
are 1, $\xi$, and $1 + \xi$, respectively. The set of the periods of $N$
prototypes is identical to the ``finger print'' ${\cal V}_0$. There is a
bijection between the set of all the periodic approximants and the set
of all the best approximants to the relevant QRQI.~\cite{EH85}
 
 So far we have confined our argument to periodic approximants of a
single QL. If an SR is applied to a periodic approximant to the QL, we
obtain a periodic approximant to another QL in the same MLD class to
which the original QL belongs.

\subsection{Duality}\label{Duality}
 The properties of a QL are decisively ruled by the relevant quadratic
irrational $\xi$. A quadratic irrational has the sole algebraic
conjugate, and to take the algebraic conjugate of a quadratic irrational
in the quadratic field ${\bf Q}[\xi]$ is an automorphism of ${\bf
Q}[\xi]$. This automorphism is a dual relationship, so that the QL may
have some properties associated with the duality.

 If the mother lattice $\Lambda$ is a square lattice, it has a canonical
set of basis vectors $\{{\bf a}, ~{\bf b}\}$ satisfying the conditions:
 \begin{equation}
    {\bf a}^2 = {\bf b}^2, \quad {\bf a}\cdot {\bf b} = 0.
   \label{orthonomal}
 \end{equation}
 If ${\bf a}$ and ${\bf b}$ are represented as Eq.\ (\ref{bases}), this
condition leads $aa_{\perp} + bb_{\perp} = 0$, which is equivalent to
$\xi = \zeta$ because of Eq.\ (\ref{ratio}). Since $\zeta = {\tilde
\xi}$, this means that $\xi$ is strongly self-dual. Conversely, if $\xi$
is strongly self-dual, the condition (\ref{orthonomal}) is satisfied
provided that the scale of the internal space is chosen
appropriately. It follows that the mother lattice of an MLD class of
binary QLs is taken to be a square lattice if and only if the relevant
MEC of the normal UFMs is strongly self-dual.

 If $\xi$ is an RQI, so is its dual ${\tilde \xi}$. We shall denote the
mother lattices associated with the two RQIs as $\Lambda$ and ${\tilde
\Lambda}$, respectively; the two are dual to each other. Also, ${\cal
W}$ and ${\cal V}$ (or ${\cal W}_0$ and ${\cal V}_0$) are dual to each
other. Since the projection modules of the two mother lattices are given
by $\Lambda_{\parallel} = {\bf Z}[\xi]$ and ${\tilde \Lambda} = {\bf
Z}[{\tilde \xi}]$, the two mother lattices coincide if and only if the
relevant MEC of the normal UFMs is self-dual; we may say then that
$\Lambda$ is self-dual. If $\Lambda$ is self-dual, we obtain ${\cal W} =
{\cal V}$ and ${\cal W}_0 = {\cal V}_0$. We will discuss below the role
of the dual mother lattice ${\tilde \Lambda}$.
 
 Let $V$ be a finite interval in the physical space, and let $Q(V, ~W) :
= Q(W) \cap V$. Then, $Q(V, ~W)$ is a projection of $\Lambda \cap R$
onto the physical space, where $R := V \times W$, the direct product, is
the rectangle obtained as a cut of the strip used to derive $Q(W)$ in
the projection method; we may write $Q(W) = Q({\bf R}, ~W)$ with ${\bf
R}$ being the entire physical space. Let ${\tilde Q}(V, ~W)$ be the
projection of $\Lambda \cap R$ onto the internal space. Then, we have
${\tilde Q}(V, ~W) = {\tilde Q}(V) \cap W$ with ${\tilde Q}(V) :=
{\tilde Q}(V, ~ {\bf R})$, where ${\bf R}$ is the entire internal space. 
It is evident that ${\tilde Q}(V)$ is a 1D QL in the internal space, and
the mother lattice of ${\tilde Q}(V)$ is ${\tilde \Lambda}$. We may say
that $Q(W)$ and ${\tilde Q}(V)$ (or, also, $Q(V, ~W)$ and ${\tilde Q}(V,
~W)$) are dual to each other.
 
 The QL ${\tilde Q}(V)$ is binary, if and only if $V$ is a canonical
interval. The two QLs in the dual pair can be mutually similar if and
only if $\Lambda$ is self-dual. At any rate, the dual relation between
RQIs is associated with a symmetry between the physical space and the
internal space. Note that $Q(V, ~W)$ and ${\tilde Q}(V, ~W)$ have a
common number of lattice points. Moreover, they can be building blocks
of the relevant periodic approximants. The present discussion on the
dual relation applies to QLs in higher dimensions as well. The mother
lattices of octagonal, decagonal, and dodecagonal QLs in 2D are shown to
be all self-dual and so are for the cases of the three types (i.e., $P$,
$I$, and $F$) of icosahedral QLs in 3D.
 
 The segments $Q(V, ~W)$ with different canonical pairs $\{V, ~W\}$ are
not necessarily independent because the hyper-scaling symmetry of
$\Lambda$ yields $Q(V, ~W) = {\tau}^{n}Q({\tau}^{-n}V, ~{\tau}^{n}W)$
for ${}^\forall n \in {\bf Z}$. Similarly we have ${\tilde Q}(V, ~W)=
{\tau}^{n}{\tilde Q}({\tau}^{n}V, ~{\tau}^{-n}W)$. Note that the area of
the rectangle, $R := V \times W$, is invariant against the
hyper-scaling.

 The structure factor of a QL, $Q$, is derived from the Fourier
transform $Q^{*}$ of the QL, where $Q$ and $Q^{*}$ are treated as Dirac
measures (generalized functions).~\cite{cite9} Let $\Lambda^{*}$ be the
reciprocal lattice to $\Lambda$. Then, the support of $Q^{*}$ is the
Fourier module, which is the projection of $\Lambda^{*}$ onto the
reciprocal physical space $E_{\parallel}^{*}$. Let $A := ({\bf a} \; ~
{\bf b})$ be a 2D matrix formed of the two basis vectors of $\Lambda$
and $A^{*} := ({\bf a}^{*} \; ~{\bf b}^{*})$ the counterpart in the
reciprocal space. Then, we may write
\begin{eqnarray}
   A = \left( \begin{array}{cc}1 & \xi \\ -\zeta & 1 \\
          \end{array}\right), \quad
   A^{*} = a^{*} \left(\begin{array}{cc}1 & \zeta \\ -\xi & 1 \\
          \end{array}\right) \label{eqn:mld20b}
\end{eqnarray}
 with $a^{*} := 2\pi/(1 + \xi\zeta)$ because $({}^{\rm t}\!A^{*})A =
 2\pi I$. Thus, $\Lambda$ and $\Lambda^{*}$ are dual to each other
 except a scale factor. The effect of the hyper-scaling on to
 $\Lambda^{*}$ is represented as $TA^{*} = A^{*}\,{}^{\rm t}\!M$. The
 structure factor of $Q$ is closely related to a 1D QL whose mother
 lattice is $\Lambda^{*}$. As a consequence, it is approximately
 self-similar. If $\Lambda$ is self-dual in particular, the structure
 factor has a similar structure to the QL itself. Note, however, that
 $\Lambda$ and $\Lambda^{*}$ are not dual for the case of icosahedral
 QLs of types $I$ and $F$.~\cite{comment2}

\section{Summary and Discussions}\label{sec_discussions}

 We have investigated 1D binary QLs with geometrical self-similarities,
and found that there exists a bijection between any pair of the four
sets: i) the set of all the MLD classes of binary QLs, ii) the set of
all the mother lattices with hyperscaling symmetries, iii) the set of
all the MECs of quasi-normal UFMs, and vi) the set of all the MECs of
QRQIs. If a mother lattice with a hyperscaling symmetry is fixed, the
following the six $N$-cycles are isomorphic to one another: i) the
relevant set of all the MLD classes of binary QLs, ii) the relevant MEC
of quasi-normal UFMs, iii) the relevant MEC of QRQIs, iv) the set of all
the scaling-equivalent class of the canonical basis vectors, v) a
similar set of the canonical windows, and vi) a similar set of the
canonical intervals. The algorithm which determines the structure of the
cycle has been independently given for the three cycles, ii), iii), and
iv), although the three algorithms are equivalent. The algorithm which
determines the structure of the cycle iii) is identical to the algorithm
for the continued-fraction expansion of a QRQI, $\xi$. It is equivalent
to the anti-diagonal algorithm which determines the structure of the
cycle iv), while the diagonal algorithm is identical to the algorithm
for the continued-fraction expansion of a QRQI, $\zeta$.  Two successive
members of every MLD class of binary QLs are combined by one of the two
elementary SRs, and arbitrary chosen two members are combined by
composite SRs. The IR of a QL is given as a special composite SR.
  
 Every 1D binary QL has a symmetric IR. A necessary and sufficient
condition for an SR to be symmetric is that its substitution matrix is a
quasi-permutation-matrix (QPM). A conventional MLD class can be divided
into two or more s-MLD classes. There exists a bijection between the set
of all the s-MLD classes of binary QLs and the set of all the MECs of
QPMs. Each s-MLD class is a cycle, and its two successive members are
combined by elementary symmetric SRs.
  
 There exist two quadratic irrationals, $\xi$ and $\tau$, characterizing
an MLD class of binary QLs. Of the two, $\xi$ includes complete
information on the MLD class but $\tau$ does not in general, so that the
former is more fundamental. The latter has complete information only
when the case of MLD classes of rank one or of a special series of MLD
classes of rank two.
  
 The binary QLs investigated in the present paper have IRs whose
inflation matrices are unimodular. Moreover, the IRs are compositions of
the elementary SRs of the two types given by Eqs.\ (\ref{ruleS}) and
(\ref{ruleU}). Then, general arguments given in Refs.\
\onlinecite{cite7} and \ \onlinecite{cite8} result in that the relevant
trace maps have the invariant of the standard form: $I := x^2 + y^2 +
z^2 - 2xyx - 1$. Therefore, the energy spectra and the one-electron wave
functions of the tight-binding models on these structures are fractals
which belong to the same classes as those of the Fibonacci
lattice.~\cite{KKT83}. Furthermore, we can conclude that each MLD class
of binary QLs has its own universality class with respect to the
one-electron properties because two successive members of every MLD
class are combined by one of the two elementary SRs.~\cite{cite7,cite8}.
  
 We have shown that every MLD class is coded by a cycle of natural
numbers. However, the MLD class is an $N$-cycle, and is not isomorphic
to the cycle coding the MLD class.  Therefore, we shall pursue a more
convenient coding system. Let us take the MLD class $\langle\langle
3,~2,~4 \rangle\rangle$ as an example, and transform it into the binary
code 100010100, which is a reversed concatenation of the three numbers
100, 10, and 1000 with three, two, and four digits, respectively. If the
digit 1 in the binary code is replaced by the elementary UFM $U$ and the
digit 0 by $S$, we obtain a quasi-normal (exactly normal) UFM which is
the inflation matrix of a member of the MLD class. The inflation
matrices of other members of the MLD class are obtained similarly from
cyclic permutations of the binary code. Therefore, for every binary QL,
we shall associate a cyclic binary decimal in the interval $]0, ~1]$ so
that the first period codes the relevant inflation matrix. Since we can
naturally define a cyclic equivalence between two cyclic decimals, we
can conclude that i) there exists a bijection between the set of all the
binary QLs and the set of all the cyclic binary decimals in the interval
$]0, ~1]$, ii) there exists a bijection between the set of all the MLD
classes and the set of all the cyclic equivalence classes of cyclic
binary decimals, and iii) the two cycles related in this bijection is
isomorphic. Thus, an MLD class of order $N$ is coded by an irreducible
cycle of $N$ binary digits like $\langle\langle 100010100
\rangle\rangle_{\rm b}$, where the suffix b stands for the symbol being
a binary code.

\appendix
\section{Mathematical Glossary}\label{glossary}

 Most results to be presented in this appendix are found in the
literature, e.g., Ref.\ \onlinecite{Davenport} but several results in
appendices \ref{Frobenius}, \ref{quasi-normal} and \ref{QRQI} seem new.

\subsection{Unimodular matrix}\label{unimodular matrix}

A matrix $M$ whose matrix elements are all integers is called unimodular
if $\det M=e=\pm 1$. Let $\{{\bf a},~{\bf b}\}$ a set of basis vectors
of a 2D lattice and $\{{\bf a}',~{\bf b}'\}$ be another set. Then, the
two sets are related to each other by a unimodular matrix
\begin{eqnarray}
  M= \left(
  \begin{array}{cc}
   p & q \\ r & s
  \end{array}
  \right) \label{unimodular1}
 \end{eqnarray}
  as
 \begin{equation}
 ({\bf a}' \; ~{\bf b}') = ({\bf a} \; ~{\bf b})M. \label{unimodular2}
 \end{equation}
 Conversely, if $M$ is unimodular, ${\bf a}'$ and ${\bf b}'$ form a set
of basis vectors of the 2D lattice.
 
 There exists a linear transformation $T$ so that ${\bf a}' := T{\bf a}$
and ${\bf b}' := T{\bf b}$; $T$ is an automorphism of the 2D
lattice. Since the signs of basis vectors of a lattice are not very
important, we can assume that $m = \mbox{Tr}\, M$ is a nonnegative
integer. The eigenvalues of $M$ are the roots of the quadratic equation
$z^{2}-m z +e=0$. There can be three cases, elliptic, parabolic, and
hyperbolic, depending on whether the discriminant $D := m^{2}-4 e$ is
negative, zero, or positive, respectively. We are interested in the
hyperbolic case, to which a general argument to be given hereafter will
be confined.  Then, $m$ must obey the condition $m \ge 1$ if $e = -1$
but the condition $m \ge 3$ if $e = +1$.  Of the two roots of the
quadratic equation, the one which is larger than unity is given by the
real quadratic irrational:
 \begin{equation}
  \tau = \frac{1}{2}\left(m+ \sqrt{m^{2}-4 e}
  \,\right). \label{unimodular3}
 \end{equation}
 The second root is given by $\bar{\tau}$, the algebraic conjugate of
$\tau$, and $\tau$ and $\bar{\tau}$ satisfy
 \begin{equation}
    \tau \bar{\tau}=e, \quad \tau + \bar{\tau}=m, \label{unimodular4}
 \end{equation}
 so that $|\bar{\tau}| < 1$. An important ${\bf Z}$-module is generated
by $\tau$:
 \begin{equation}
    {\bf Z}[\tau] := \{t + u\tau \,|\, t, ~u \in {\bf Z}\}.
   \label{Z-module}
 \end{equation}
 Remember that the Cayley-Hamilton theorem yields $M^2 - mM + eI = 0$
with $I$ being the unit matrix.

 The linear transformation $T$ has two principal axes which we can
assume as shown shortly to be mutually orthogonal; $T$ enlarges along
one of the axes by $\tau$ but shrinks along the other by
${\tau}^{-1}=|\bar{\tau}|$. If the coordinate system matches the
principal axes, $T$ is represented by the diagonal matrix
  \begin{equation}
  T = \left( \begin{array}{cc} \tau & 0 \\ 0 & \bar{\tau}
 \end{array} \right). \label{unimodular5}
 \end{equation}
 Let $A := ({\bf a} \; ~{\bf b})$ be a 2D matrix formed of the two
column vectors represented in this coordinate system. Then, Eq.\
(\ref{unimodular2}) is equivalent to
 \begin{equation}
  TA = AM, \label{unimodular6}
 \end{equation}
 so that $M$ is diagonalized by $A$: $AMA^{-1} = T$, and the two row
vectors of $A$ are the left eigenvectors corresponding to the two
eigenvalues, $\tau$ and $\bar{\tau}$, respectively. Therefore, the basis
vectors ${\bf a}$ and ${\bf b}$ are almost determined from $M$ by the
orthogonality of the principal axes; we have a freedom of choosing
proportionality constants of the two left eigenvectors.
 
 Let $(a \;\; b)$ be the left eigenvector given by the first row of
$A$.~\cite{comment1} Then, it satisfies the equation,
 \begin{equation}
 \tau(a \;\; b) = (a \;\; b)M. \label{unimodular7}
 \end{equation}
 Hence the ratio $\xi := b/a$ satisfies $\tau (1 \;\; \xi)=(1 \;\;
\xi)M$. It follows that
 \begin{equation}
    \tau = p + r\xi, \quad \tau\xi = q + s\xi. \label{unimodular8}
 \end{equation}
 Hence $\xi$ is a real quadratic irrational belonging to the quadratic
field ${\bf Q}[\tau]$. It is important that the ${\bf Z}$-module ${\bf
Z}[\xi]$ coincides with ${\bf Z}[\tau]$ only when $r = 1$ because ${\bf
Z}[\tau] = r{\bf Z}[\xi]$. Note that $\xi$ is a root of the quadratic
equation:
 \begin{equation}
    r\xi^2 + (p - s)\xi - q = 0. \label{unimodular9}
 \end{equation}
 In this paper, we always mean by a quadratic irrational a real
quadratic irrational.

 The ratio $\xi$ is uniquely determined by $M$ but the converse is not
necessarily true because $M^n$ with $n$ being any nonzero integer
satisfies the condition. A unimodular matrix is called irreducible, if
it cannot be written as a power of another unimodular matrix. Every
quadratic irrational $\xi$ has its proper irreducible unimodular matrix;
we shall say that $\xi$ and the relevant matrix are {\it companions} of
each other.

 Another quadratic irrational is defined with $\xi$ by ${\tilde \xi} := -
\bar{\xi}^{-1}$. We shall call it the {\it dual} to $\xi$ because the
tilde operation is recursive. The companion matrix of ${\tilde \xi}$ is
equal to ${}^{\rm t}\!M$, the transposed matrix of $M$.

The set, $GL(2, {\bf Z})$, of all the 2D unimodular matrices form an
infinite but discrete group. The set, $GL'(2, {\bf Z})$, of all the
hyperbolic and irreducible members of $GL(2, {\bf Z})$ is, however, not
a group because it does not include the unit matrix. We may say
generally that two unimodular matrices $M$ and $M'$ are {\it modular
equivalent} if there exists a unimodular matrix $K$ so that $M' =
K^{-1}MK$. The set $GL'(2, {\bf Z})$ can be grouped into modular
equivalent classes (MECs), which are disjoint. The two eigenvalues are
common among different members of an MEC. Also the trace and the
determinant are common, so that they are proper numbers of the
class. Note, however, that the pair of the two numbers is not sufficient
to specify an MEC. For example, the two matrices, $\left({0 \atop 1}\;
{1 \atop 6}\right)$ and $\left({2 \atop 3}\; {3 \atop 4}\right)$ are
shown to be not modular equivalent.

There exists a bijection (one-to-one correspondence) between any pair of
the three sets: i) $GL(2, {\bf Z})$, ii) the set of all the sets of
basis vectors of a 2D lattice, and iii) the set of all the automorphisms
of the 2D lattice. Therefore, we can define primed subsets like $GL'(2,
{\bf Z})$ for the latter two sets as well, and the division of $GL'(2,
{\bf Z})$ into MECs induces similar divisions of other two primed sets.

\subsection{Unimodular Frobenius matrices}\label{Frobenius}

 The case where $M \in GL'(2, {\bf Z})$ is a Frobenius matrix, whose
matrix elements are all nonnegative, is of particular importance. The
positive eigenvalue, $\tau$, of $M$ is called the Frobenius eigenvalue,
and the relevant left or right eigenvector is called the Frobenius
eigenvector, which is assumed to have two positive components. It
follows that the companion quadratic irrational $\xi$ is positive.

 The set of all the Frobenius matrices in $GL(2, {\bf Z})$ do not form a
group but only do a semigroup because the inverse of a {\it unimodular
Frobenius matrix} (UFM) has not necessarily the Frobenius property. Let
$M$ and $M'$ be two UFMs which are modular equivalent to each
other. Then, we say that they are {\it strongly modular equivalent} to
each other if there exist two UFMs $K$ and $L$ so that $M' = K^{-1}MK$
and $M = L^{-1}M'L$.
This condition is equivalent to the one that $M$ is a
product of the two UFMs, $M = KL$. More generally, if $M$ is a product
of a number of UFMs, it is strongly modular equivalent to another UFM
which is obtained from $M$ by a cyclic permutation of its factors in the 
product. 

 A UFM represented as Eq.\ (\ref{unimodular1}) is called {\it
 quasi-normal} if it satisfies the two conditions: i) it is irreducible
 and ii) $p \le q$ and $r \le s$.
 It is called {\it normal} if it satisfies the condition, iii) $p \le r$
 and $q \le s$, in addition to the former two. A quasi-normal UFM is
 normal if and only if its transpose is also quasi-normal.
 In particular, a UFM and its transpose are simultaneously normal. Note
 that a normal UFM is always hyperbolic but a quasi-normal UFM can be
 parabolic. 

 The set of all the quasi-normal (or normal) UFMs in $GL(2, {\bf Z})$
form a semigroup if the unit matrix is included in it.
If an irreducible UFM $M$ ($\ne I$) is not quasi-normal, $P^{-1}MP$ with
$P := \left({0 \atop 1}\; {1 \atop 0}\right)$ is quasi-normal. 
That is, a UFM being not
quasi-normal is strongly modular equivalent to a quasi-normal
UFM. Therefore, properties of the former are investigated from those of
the latter. An important property of a quasi-normal UFM is that its
companion quadratic irrational is larger than unity, which follows from
Eq.\ (\ref{unimodular8}).

 Simplest but nontrivial quasi-normal UFMs are:
 \begin{eqnarray}
   S = \left( \begin{array}{cc} 1 & 1 \\ 0 & 1 \\
          \end{array}\right), \quad
   U = \left( \begin{array}{cc} 0 & 1 \\ 1 & 1 \\
          \end{array}\right). \label{SUM}
 \end{eqnarray}
 In fact, the second of the two is normal. A series of normal UFMs is
 defined by
 \begin{eqnarray}
   K := \left(
    \begin{array}{cc}
     0 & 1 \\ 1 & k
    \end{array}
  \right) \label{K}
 \end{eqnarray}
 with $k$ being natural numbers. We shall call a member of the series an
{\it elementary normal UFM}. The matrix (\ref{K}) is factorized as $K =
US^{k-1}$.  Similarly, a series of quasi-normal UFMs is defined by
 \begin{equation}
  L= \left(
   \begin{array}{cc}
    1 & l \\ 0 & 1
   \end{array}
   \right) \label{L}
 \end{equation}
 with $l$ being natural numbers. We may write $L = S^l$, which is a
generic form of a parabolic quasi-normal UFM.

 If a normal UFM is represented as Eq.\ (\ref{unimodular1}), we obtain
 \begin{equation}
  MK^{-1}= \left(
   \begin{array}{cc}
   q - kp & p \\ s - kr & r
   \end{array}
   \right). \label{LM}
 \end{equation}
 We shall denote by $k_0$ the maximum number $k$ under the condition
that the unimodular matrix $MK^{-1}$ remains a UFM. Then, $MK^{-1}_0$ is
a normal UFM unless it is equal to the unit matrix. That is, a normal
UFM is transformed into another normal UFM with a smaller ``norm''
provided that the norm is defined appropriately. We can repeat this
procedure until we arrive at an elementary normal UFM. Thus, a normal
UFM is uniquely factorized into elementary normal UFMs:
 \begin{equation}
   M = K_{n-1}K_{n-2} \cdots K_{0}. \label{frobenius2}
 \end{equation}
 The {\it rank} of $M$ is defined by the total number $n$ of the factors
in the right hand side of this expression. We should mention that the
algorithm of this factorization is nothing but a parallel Euclidean
algorithm. We can specify $M$ by the ordered set $\langle k_{0},~k_{1}, ~
\cdots, ~k_{n-1} \rangle$ of natural numbers, where $k_{i}$ is a number
specifying $K_{i}$; we have adopted the reversed order for a convenience
of a later argument. The members of the ordered set are not entirely
independent because of the irreducibility condition i) above. For
example, $k_{0}$ and $k_{1}$ must be different if $n = 2$. If $M =
\langle k_{0},~k_{1}, ~\cdots, ~k_{n-1} \rangle$, then ${}^{\rm t}\!M =
\langle k_{n-1},~k_{n-2}, ~\cdots, ~k_{0} \rangle$; $M$ is symmetric if
and only if the relevant ordered set is mirror-symmetric.

 As a result of the uniqueness of the factorization above, we can
conclude that a normal UFM which is modular equivalent to the normal UFM
(\ref{frobenius2}) has a similar factorization to above but the factors
are cyclically permuted. Hence, the rank is common between the two UFMs. 
An MEC of normal UFMs is composed of $n$ members, where $n$ is the
common rank of them. The MEC is specified by the cyclically ordered set
$\langle\langle k_{0},~k_{1}, ~\cdots, ~k_{n-1} \rangle\rangle$, where
double angular brackets are used to distinguish the set from the
linearly ordered set above. Note that the sign of the determinant of
every member of the MEC is equal to $(-1)^n$. We will call hereafter a
cyclically ordered set simply a {\it cycle}.
 
 An MEC is called {\it self-dual} if it is specified by a cycle with a
mirror symmetry. Every member of a self-dual MEC is modular equivalent
to its transpose. An MEC is always self-dual if its rank is less than
three. An MEC of rank three is self-dual if and only if it takes the
form $\langle\langle k,~k', ~k \rangle\rangle$, while an MEC of rank
four is self-dual if and only if its form is one of the two
alternatives, $\langle\langle k,~k',~k', ~k \rangle\rangle$ and
$\langle\langle k,~k',~k'', ~k' \rangle\rangle$.
 
  An MEC is called {\it strongly self-dual} if it includes a symmetric
member (i.e., a symmetric matrix). A self-dual MEC is strongly self-dual
if its rank is odd but not necessary so if its rank is even. For
example, an MEC of rank two is never strongly self-dual, while an MEC of
rank four is strongly self-dual if and only if its form is the former of
the two alternatives above.

\subsection{Quasi-normal UFM and quasi-permutation-matrix}\label{quasi-normal}

 Let $M'$ be a quasi-normal UFM. Then, we can show by a similar argument
to that presented after Eq.\ (\ref{LM}) that there exists the maximum
number $l$ under the condition that $L^{-1}M'$ remains a UFM. In the
maximal case, $L^{-1}M'$ is a normal UFM. Then, we can show that $M : =
L^{-1}M'L$ is a normal UFM as well, so that a quasi-normal UFM is
strongly modular equivalent to a normal UFM. It follows that a
quasi-normal UFM is represented with a normal UFM $M$ as $ S^{l}MS^{-l}$
with $l$ being a nonnegative integer. More precisely, if $M$ is
represented as Eq.\ (\ref{frobenius2}), we obtain $k_{0}$ quasi-normal
UFMs:
 \begin{equation}
   M_k = S^{k}U S^{k_{n-1}-1}U S^{k_{n-2}-1} \cdots
  US^{k_{1}-1}US^{k_{0}-k-1} \label{M'}
 \end{equation}
with $k = 0, ~1, ~2, ~\cdots, ~k_{0} -1$ and $M_0 := M$, where use has
been made of the equations, $K_{i}= US^{k_{i}-1}$. Note that the
factorization of $M_k$ like this is unique. More strongly, there exists
an algorithm by which each factor in such a factorization is determined
step by step: the right end factor of a quasi-normal UFM $M$ is $S$ if
$MS^{-1}$ is quasi-normal but $U$ otherwise. The two elementary UFMs,
$U$ and $S$, are two of the four which are given in Ref.\
\onlinecite{cite7} as generators of UFMs. The remaining two are not
necessary here because the UFMs under consideration are restricted to
quasi-normal ones.

If $\langle\langle k_{0},~k_{1}, ~\cdots, ~k_{n-1} \rangle\rangle$ is an
MEC of normal UFMs, we can obtain $k_{i}$ quasi-normal UFMs from the
$i$-th member of the class; the total number of quasi-normal UFMs
obtained is equal to $N :=k_{0}+k_{1}+ \cdots +k_{n-1}$. They form an
MEC of quasi-normal UFMs, which are strongly modular equivalent to one
another. The two MECs, one normal and the other quasi-normal, can be
specified by a common cycle of natural numbers although the latter MEC
is a cycle composed of $N$ members. If a member of an MEC of
quasi-normal UFMs is factorized as Eq.\ (\ref{M'}), $n$, the number of
$U$, and $N$, the total number of the factors, are proper numbers of the
MEC; $n$ is nothing but the rank of the MEC. We shall call $N$ the {\it
order} of the MEC.

The subject in the remaining part of this subsection concerns only in
Sec.\ \ref{sec:symmetric} in the text. There are only two permutation
matrices in 2D, i.e., the unit matrix $I$ and $J := \left({0 \atop 1}\;
{1 \atop 0}\right)$. We shall call a UFM a {\it
quasi-permutation-matrix} (QPM) if it is congruent in modulo 2 with $I$
or $J$. The product of two or more QPMs is also a QPM. If a UFM is not a
QPM, it is congruent in modulo 2 with one of the four matrices, $S$,
${}^{\rm t}\!S$, $U$, and $R : = \left({1 \atop 1}\; {1 \atop 0}\right)$
because $\det M = \pm 1$. In what follows, the congruence relation $x
\equiv y \bmod 2$ is simply written as $x \equiv y$. Then, the four
matrices just introduced satisfy $S^2 \equiv ({}^{\rm t}\!S)^2 \equiv
I$, $U^2 \equiv R$, $R^2 \equiv U$, and $U^3 \equiv R^3 \equiv I$.

Since $m \equiv 0$ or $m \equiv 1$ according as $m := {\rm Tr}\;M$ is
even or odd, respectively, a UFM with an odd $m$ cannot be a QPM. More
precisely, $M \equiv U$ or $M \equiv R$ for this case, so that $M^2$ is
not a QPM but $M^3$ is; $M^3 \equiv I$. On the other hand, a UFM with an
even $m$ satisfies $M \equiv S$ or $M \equiv {}^{\rm t}\!S$ if $M$ is
not a QPM. Then, $M^2 \equiv I$.
 
 A QPM is called {\it elementary}, if it cannot be represented as a
product of two QPMs. Every QPM is uniquely factorized into elementary
QPMs. The two QPMs, $S^2$ and $US$, are simplest elementary QPMs which
are quasi-normal. We present other four types of elementary and
quasi-normal QPMs: $SUS^kU \equiv I$, $SUS^lU \equiv J$, $US^lUS^{l'}U
\equiv I$, and $US^lUS^kU \equiv J$, where $k$ is an odd integer but $l$
and $l'$ are even. There can be more complicated QPMs, e.g.,
$U(US^k)^{k'}U$ with $k$ and $k'$ being any odd numbers, but we shall
not pursuit them. A QPM is called {\it irreducible}, if it cannot be
written as a power of another QPM. If a composite QPM is factorized into
elementary QPMs, we obtain another QPM by a cyclic permutation of its
factors, and the two QPMs are strongly modular equivalent to each other. 
Note that an irreducible QPM can be reducible as a usual UFM. For
example, if $M$ is not a QPM, $M^2$ or $M^3$ is an irreducible QPM.

Finally, an example of a factorization of a QPM into elementary QPMs is
presented:
 \begin{equation}
    (US^{4})^{3}= (US)(S^2)(SUS^{4}U)(S^2)^{2}. \label{cycletriple}
 \end{equation}

\subsection{Modular transformation}\label{modular transformation}
 
 Let us take a unimodular matrix
 \begin{equation}
  X= \left(
  \begin{array}{cc}
   t & u \\ v & w
  \end{array}
  \right) \in GL(2, {\bf Z}). \label{X}
 \end{equation} 
 Then, a transformation from an irrational number $\xi$ to another
irrational $\xi'$ is uniquely defined by the condition
  \begin{equation}
  (1 \;\; \xi) = \lambda(1 \;\; \xi')X \; \mbox{or} \; \lambda(1 \;\;
\xi') = (1 \;\; \xi)X^{-1}, \label{modular0}
 \end{equation}
 where $\lambda$ is a parameter determined by this condition. This
transformation is called the {\it modular transformation}, which is
written explicitly as
 \begin{equation}
  \xi'=X(\xi) :=\frac{\;\;t\xi - u}{-v\xi + w}. \label{modular1}
 \end{equation}
 Our definition of the modular transformation is slightly different from
the conventional one but is substantially equivalent to the latter. If
$\xi'=X(\xi)$ and $\xi''=X'(\xi')$ with $X'$ being the second unimodular
matrix, we obtain $\xi''=X''(\xi)$ with $X'' := X'X$, which can be
readily proved from Eq.\ (\ref{modular0}). Therefore, the set of all the
modular transformations form a group which is homomorphic with $GL(2,
{\bf Z})$. More precisely, it is isomorphic with $GL(2, {\bf Z})/Z_2$,
where $Z_2 := \{I, ~-I\}$ is a normal subgroup of $GL(2, {\bf Z})$
because two unimodular matrices $X$ and $-X$ yield an identical modular
transformation.

A quadratic irrational is changed into another quadratic irrational by a
modular transformation. Hence, the set of all the quadratic irrationals
can be divided into MECs, which are disjoint.

 Let us introduce with the unimodular matrix $X$ a new set of basis
 vectors by
 \begin{equation}
 ({\bf a}' \; ~{\bf b}') = ({\bf a} \; ~{\bf b})X. \label{modular2}
 \end{equation}
The first row of this equation reads
 \begin{equation}
 (a' \;\; b') = (a \;\; b)X. \label{modular3}
 \end{equation}
 It follows that the ratio $\xi' := b'/a'$ is related to $\xi$ by the
modular transformation, $\xi'=X^{-1}(\xi)$. As a consequence, we can
conclude that $\xi$ is a fixed point of a modular transformation
associated with its companion matrix $M$: $\xi=M(\xi)$. Moreover, we
obtain $\xi'=M'(\xi')$ with $M' := X^{-1}MX$. Thus, there exists a
bijection between the set of all the MECs of quadratic irrationals and
the set of all the MECs of (hyperbolic) unimodular matrices.
 
 The following proposition provides us with a different view for modular
equivalence between two irrationals, $\xi$ and $\xi'$: a necessary and
sufficient condition for $\xi$ and $\xi'$ to be modular equivalent is
that the two ${\bf Z}$-modules, ${\bf Z}[\xi]$ and ${\bf Z}[\xi']$, are
scaling-equivalent: ${\bf Z}[\xi'] = \mu{\bf Z}[\xi]$ for ${}^\exists
\mu \in {\bf R}$; if $\xi'=X^{-1}(\xi)$ with $X$ being the unimodular
matrix (\ref{X}), the multiplier $\mu$ is determined as $\mu = t +
v\xi$. The scaling equivalence between ${\bf Z}$-modules is essential in
a classification of quasilattices in any dimensions.~\cite{RMW}

\subsection{Number theory of continued fractions}\label{continued fractions}

 Let $\xi$ be an irrational number being larger than unity. Then, we can
derive a second irrational $\xi' > 1$ by the equation:
 \begin{equation}
  \xi = k + \frac{1}{\xi'}, \quad k := \lceil\xi\rceil,
  \label{continued-fraction1}
 \end{equation} 
 where the symbol $\lceil*\rceil$ stands for the maximum integer among
those which do not exceed the number $*$. This equation is written with
the use of the modular transformation as
 \begin{equation}
  \xi' = K(\xi), \label{continued-fraction2}
 \end{equation}
 where $K$ is the normal UFM (\ref{K}). Since $K = US^{k-1}$ with $S$
and $U$ being UFMs (\ref{SUM}), this modular transformation is
represented as a composition of the two types of elementary procedures:
  \begin{equation}
    \xi' = S(\xi) = \xi - 1, \quad \xi' = U(\xi) = \frac{1}{\xi - 1}.
\label{continued-fraction4}
 \end{equation} 
 The first of the two procedures is used when $\xi > 2$, while the
second when $1 < \xi < 2$.
 
 If the recursive procedure (\ref{continued-fraction1}) is repeated
indefinitely, $\xi$ is expanded into an infinite continued-fraction. The
$i$-th step of this recursive procedure is given by
 \begin{equation}
  \xi_{i} = k_{i} + \frac{1}{\xi_{i + 1}} \label{continued-fraction5}
 \end{equation} 
 with $\xi_{0} := \xi$ and $k_{0} := k$, so that
 \begin{equation}
  \xi_{0} = k_{0}+ \frac{1}{k_{1} + {}} \frac{1}{k_{2} + {}} \cdots
   \frac{1}{k_{i-1} + {}} \frac{1}{\xi_{i}}. \label{continued-fraction6}
 \end{equation} 
 This equation is equivalent to $\xi_{i}=M_{i}(\xi_{0})$ with $M_{i} =
K_{i-1}K_{i-2} \cdots K_{0}$ being a normal UFM, where $K_{j}$ is given
by Eq.\ (\ref{K}) with $ k = k_{j}$.  If $\xi_{i}$ in the
continued-fraction (\ref{continued-fraction6}) is replaced by an integer
$k$ satisfying $1 \le k \le k_{i}$, we obtain a rational number, which
is shown to be a ``best'' approximant to $\xi_{i}$.  ~\cite{Takagi}

 Numbers having periodic continued-fraction expansions are
important. Let $M$ be the normal UFM derived from the first period of
the expansion of such number, $\xi$. Then $\xi$ is the companion
quadratic irrational of $M$. Hence, $\xi$ must be a quadratic
irrational. More strongly, $\xi>1$ and ${\tilde \xi}>1$ because $M$ is
normal. However, the continued-fraction expansion of a generic quadratic
irrational is not always periodic; exactly, it is periodic except for
first several terms.
 
 A quadratic irrational is called {\it a reduced quadratic irrational}
(RQI) if its continued-fraction expansion is periodic. A necessary and
sufficient condition for a quadratic irrational $\theta$ to be an RQI is
that it satisfies the conditions: $\theta>1$ and ${\tilde \theta}>1$,
which is equivalent to $-1<\bar{\theta}<0$. Thus, there exists a
bijection between the set of all the normal UFMs and the set of all the
RQIs. Then, the symbols used to specify each normal UFM and each MEC of
normal UFMs are used to specify the corresponding RQI and relevant MEC
of RQIs, respectively. The rank $n$ of an RQI is nothing but the
shortest period of its continued-fraction expansion, and the number of
the members of the relevant MEC of RQIs is equal to $n$.
 
The subject in the remaining part of this subsection concerns only in
Sec.\ \ref{Duality} in the text. Let $\theta$ be an RQI specified by the
ordered set $\langle k_{0},~k_{1}, ~\cdots, ~k_{n-1} \rangle$. Then, the
dual ${\tilde \theta}$ to $\theta$ is specified by the inversely ordered
set. An MEC of RQIs is called self-dual or strongly self-dual if the
corresponding MEC of normal UFMs is self-dual or strongly self-dual,
respectively. A strongly self-dual MEC of RQIs includes an RQI $\theta$
so that ${\tilde \theta} = \theta$; the first period of the
continued-fraction expansion of $\theta$ is mirror-symmetric. We may say
this RQI to be strongly self-dual. If $\xi$ is a member of a self-dual
MEC of RQIs, ${\tilde \xi}$ is modular equivalent to $\xi$ or,
equivalently, ${\bf Z}[\xi]$ and ${\bf Z}[{\tilde \xi}]$ are
scaling-equivalent.
 
 A sufficient condition for an MEC of RQIs to be self-dual is that it
includes an RQI $\xi$ so that $\xi + {\bar \xi} \in {\bf Z}$ or,
equivalently, $\xi$ takes the form $\xi := p + q\sqrt{D}\,$ or $\xi :=
(p + q\sqrt{D}\,)/2$, where $p$, $q$, and $D$ are natural
numbers. Therefore, the one period of the continued fraction expansion
of a quadratic irrational of either of the two forms is mirror symmetric
if it is considered to be a cycle.

\subsection{Quasi-reduced quadratic irrational}\label{QRQI}

 The companion quadratic irrational $\xi$ of a quasi-normal UFM is a
{\it quasi-reduced quadratic irrational} (QRQI) which is defined in
Refs.\ \onlinecite{s2,s3} because it satisfies the conditions: $\xi>1$
and ${\tilde \xi}>0$ (or, equivalently, $\bar{\xi}<0$). A necessary
condition for a quadratic irrational being larger than unity to be an
QRQI is that its continued-fraction expansion is periodic but for the
first term.
It can be shown readily that a QRQI $\xi$ is related to an RQI
$\theta$ as $\xi=\theta-k$ with $k$ being an integer satisfying the
inequality, $0 \le k \le k_{0} -1$ with $k_{0} := \lceil\theta\rceil$.

The relation $\xi=\theta-k$ between a QRQI $\xi$ and an RQI $\theta$ is
written in terms of the modular transformation specified by a UFM $S$
given in Eqs.\ (\ref{SUM}) as $\xi= S^{k} (\theta)$. It follows from the
equation $\theta= M_0 (\theta)$ with $M_0$ being the companion matrix of
$\theta$ that $\xi$ is the companion quadratic irrational of the
quasi-normal UFM, $M_k= S^{k}M_0S^{-k}$: $\xi=M_k(\xi)$. An explicit
form of $M_k$ is given by Eq.\ (\ref{M'}), where we have set $M_0 : =
\langle k_{0},~k_{1}, ~\cdots, ~k_{n-1} \rangle$. The RQI $\theta$
yields $k_{0}$ QRQIs, $\xi=\theta-k$ with $k = 0, ~1, ~2, ~\cdots, ~
k_{0} -1$, which we shall call {\it companion QRQIs} of $\theta$. Thus,
there exists a bijection between the set of all the QRQIs and the set of
all the quasi-normal UFMs. More strongly, there exists a bijection
between the set of all the MECs of QRQIs and that of the quasi-normal
UFMs. Both the two types of MECs combined by the bijection have
structures of $N$-cycles, and these structures are retained in the
bijection. That is, the two $N$-cycles are isomorphic. Therefore, we can
specify an MEC of QRQIs by the symbol specifying the corresponding MEC
of quasi-normal UFMs: $\langle\langle k_{0},~k_{1}, ~\cdots, ~k_{n-1}
\rangle\rangle$ with $n$ being the rank of the class. Thus, the relevant
MEC of QRQIs includes $n$ RQIs, $\theta_j$ with $j = 0, ~1, ~2, ~\cdots,
~n -1$ and $\lceil\theta_j\rceil= k_{j}$. The companion QRQIs of
$\theta_j$ are $\xi^{(j)}_k :=\theta_j-k$ with $k = 0, ~1, ~2, ~\cdots, ~
k_{j} -1$.

 As a typical example of the MEC we take $\langle\langle 1,~2,~1
\rangle\rangle$, which is composed of the four QRQIs:
 \begin{eqnarray*}
   \xi^{(0)}_{0} &=& \frac{1+\sqrt{10}}{3}, \quad \xi^{(1)}_{0} =
   1+\frac{\sqrt{10}}{2}, \\ \xi^{(1)}_{1} &=& \frac{\sqrt{10}}{2},
   \quad \xi^{(2)}_{0} = \frac{2+\sqrt{10}}{3}.
 \end{eqnarray*}
 The ratio $\tau$ associated with this MEC is given by
 $\tau=3+\sqrt{10}\,$. The companion matrix of $\xi^{(0)}_{0}$, for
 example, is given by
 \begin{equation}
  M= \left(
   \begin{array}{cc}
    2 & 3 \\ 3 & 4
   \end{array}
   \right). \label{inflmtrx2334}
 \end{equation}
 This example can be used to check or understand various results
 appearing in the present paper.

\pagebreak

\pagebreak
 
\pagebreak
 
\pagebreak
 
\begin{references}
\bibitem{comment4} The abbreviation ``MLD'' will be used not only for
	 ``mutual local-derivability'' but also for ``mutually
	 locally-derivable''.
\bibitem{comment5} A necessary condition for two QLs to be MLD from
	 each other is that their {\it limit translation modules} are
	 the same except for their scales.~\cite{cite2}
\bibitem{comment3} The generalized Fibonacci numbers $F_n$ are
	 solutions of the recursion relation, $F_{n+1} = mF_n - eF_{n-1}$,
	 with the initial conditions, $F_0 = 0$ and $F_1 = 1$, and the
	 ratio $F_{n+1}/F_n$ tends to $\tau$ as $n$ goes to the
	 infinity. Alternatively, $F_n$ are defined by
	 $\tau^n = F_n\tau - eF_{n-1}$.
\bibitem{comment2} Note that $\Lambda$ and $\Lambda^{*}$ are dual by
	 the very definition in the linear algebra. The word ``dual''
	 defined in this paper is number-theoretical, and is indifferent
	 to that definition.
\bibitem{comment1} In this paper, a 2D vector $(a, ~b)$ will be denoted
	 as $(a \;\; b)$ if it is regarded as a row vector (i.e., a
	 $1\times 2$ matrix).
\end{references}
\end{document}